\documentclass[a4paper,11pt]{article}
\pdfoutput=1 

\usepackage{jcappub} 

\usepackage[T1]{fontenc} 
\usepackage{relsize}
\usepackage{booktabs}

\newcommand{\ra}[1]{\renewcommand{\arraystretch}{#1}}
\def\be{\begin{equation}}
\def\ee{\end{equation}}
\def\ba{\begin{eqnarray}}
\def\ea{\end{eqnarray}}
\def\bs{\begin{split}}
\def\es{\end{split}}
 
\title{Disformal couplings and the dark sector of the universe}

\author{C. van de Bruck}
\author{and J. Morrice}
\affiliation{Consortium for Fundamental Physics,\\
	School of Mathematics and Statistics,\\
	University of Sheffield,\\
	Hounsfield Road,\\
	Sheffield S3 7RH, United Kingdom.}

\emailAdd{c.vandebruck@sheffield.ac.uk}
\emailAdd{app12jam@sheffield.ac.uk}

\abstract{Interactions between dark matter and dark energy, allowing both conformal and and disformal couplings, are studied in detail.  We discuss the background evolution, anisotropies in the cosmic microwave background and large scale structures.
One of our main findings is that a large conformal coupling is not necessarily disallowed in the presence of a general disformal term. On the other 
hand, we find that negative disformal couplings very often lead to instabilities in the scalar field. Studying the background evolution and linear 
perturbations only, our results show that it is observationally challenging to disentangle disformal from purely conformal couplings.}

\begin{document}
\maketitle
\flushbottom

\section{Introduction}
Observations of the cosmic microwave background radiation (CMB) and large scale structures (LSS) have allowed cosmologists to formulate a model of cosmology 
in which the standard model (SM) particles are a subdominant matter form. The model predicts the existence of dark 
matter which only interacts very weakly with itself and the other matter particles. In addition, the model requires an energy form with negative pressure, 
dubbed dark energy, which is responsible for the accelerated expansion of the universe in the present epoch (see \cite{Ade:2013zuv} for the 2013 results of the Planck mission). A major task of present day cosmology is to illuminate the properties of dark matter and dark energy. 

A simple candidate for dark energy is the cosmological constant. Its biggest drawback however, is that its observed magnitude is $10^{120}$ times smaller than 
the value expected from theory, when interpreted as a vacuum energy density. Because of this, cosmologists have studied other possibilities, such as dynamical scalar fields, or modified theories of gravity. We refer to \cite{Copeland:2006wr}, \cite{Amendolabook} and \cite{Clifton:2011jh} for recent reviews. Here, we focus on a union of the two: the case of a scalar field as a dark energy candidate, which modifies the force of gravity.  
In such models, couplings to all matter/energy forms are expected unless symmetries exist which forbid or suppress interactions, yet, problematically, a scalar field coupled to matter would mediate a long range fifth force between the different particles, a force which is not observed in nature \cite{Carroll:1998zi}. Such non-detection implies that the coupling to baryons must be very small, whereas constraints on coupling to neutrinos and dark matter are substantially weaker, and must be obtained from cosmological observations. And, very recently, evidence has emerged to suggest that an interaction between elements of the dark sector is not just plausible, but actually favored by current data (\cite{Salvatelli:2014zta} and \cite{Abdalla:2014cla}). The analysis was in each case purely phenomenological, assuming a minimal amount of underlying theory, yet it is a progressive step toward understanding the nature of these invisible elements of our universe.

In light of these facts, we dedicate this work to the investigation of dark energy as a very light scalar field coupled to dark matter only, and assume all interactions between the standard model and the dark sector are negligible. Theories with an interacting dark sector have been discussed in the literature extensively, and our ignorance of this sector's physical nature is reflected in the wide variety of interaction types considered;  see e.g. \cite{Wetterich:1987fm} - \cite{Amendola:2014kwa} and references therein. The setup in which the interactions are only in the dark sector has been motivated from theories with extra--dimensions and branes \cite{Koivisto:2013fta}. As such, the couplings between dark matter and dark energy are purely geometric in origin in which dark matter lives on a brane distinct from the brane on which the standard model particles are confined. In many of these works, the gravity sector of the theory is of scalar--tensor form, the scalar plays the role of dark energy, and the coupling of the scalar field to dark matter is described via a conformal transformation of its geometry - dark matter now responds to curvature via an effective Newton's constant that depends on the local value of the scalar field. The interaction is hence termed a conformal coupling.

As an extension of this idea, we now allow the additional possibility of disformal couplings between the two dark elements. Disformal models of gravity, initiated by Bekenstein \cite{Bekenstein:1992pj} have been attracting much attention recently, particularly with regards to cosmology, see \cite{Koivisto:2013fta} and \cite{Koivisto:2008ak} - \cite{Sakstein:2014aca}. These disformal factors have been used in stabilizing scaling solutions in massive gravity \cite{Mukohyama:2014rca}, modifying the speed of gravitational wave propagation during inflation \cite{Creminelli:2014wna}, even describing electron transport theory in strained graphene \cite{Juan:2012prl}, and many other ways besides. One of the central issues we address in this paper is whether or not cosmological observations will allow us to disentangle the effects of conformal and disformal couplings. As we couple the scalar field to dark matter only, we avoid the stringent constraints on disformal couplings from a host of local tests \cite{Brax:2014vva}. 

To demonstrate clearly what we mean by conformal and disformal transformations of the dark matter geometry, let us now write down the action for the theory we consider: 
\be
\label{eq:action}
\mathcal{S} = \int d^4x\sqrt{-g}\left\{ \frac{\mathcal{R}(g)}{2\kappa} + \mathcal{L}_{\rm (SM)} + \mathcal{L}_{(\rm DE)} \right\} + \int d^4x \sqrt{-\tilde{g}}\tilde{\mathcal{L}}_{(\rm DM)},
\ee
where SM corresponds to the visible sector (i.e. the standard model particles) and dark energy, $\mathcal{L}_{(DE)}$, is described by a quintessence field: 
\be
\mathcal{L}_{(DE)} = -\frac{1}{2}\nabla^{\alpha}\phi\nabla_{\alpha}\phi - V(\phi). 
\ee
The dark matter sector, described by the Lagrangian
\be
\tilde{\mathcal{L}}_{(DM)} = \tilde{\mathcal{L}}_{(DM)}(\tilde{g}_{\alpha\beta};\varphi),
\ee
depends on the metric 
\be
\label{eq:dismetric}
\tilde{g}_{\mu\nu} = C(\phi)g_{\mu\nu} + D(\phi)\phi,_{\mu}\phi,_{\nu}.
\ee
The functions $V$, $C$ and $D$ encapsulate our theory's remaining freedom, that will be specified in later sections. $C$ and $D$ go by the names `conformal factor' and `disformal factor' respectively. We see now that dark matter particles follow geodesics determined by $\tilde{g}_{\mu\nu}$, and that various aspects of these particles, for instance their mass, will now depend on the dark energy field. The functions $C$ and $D$ could also depend on the derivatives of $\phi$, but we will ignore this possibility for simplicity in this paper. 

The theory we have just made concrete is a mathematical realization of generalized gravitational interactions within the dark sector. It encompasses a very broad, though not exhaustive, number of alternatives to general relativity, which is now just a point in the function space: $C=1$ and $D=0$. The case of $D=0$ but $C\neq1$ has been extensively discussed, and only recently have cosmologists studied the implications of the disformal term as well. In this paper we will be studying models with $C\neq 1$ and $D\neq 0$ and compare them to the case of purely conformal ones. In particular, the interaction between the two types is investigated numerically, and we will find this interplay has important consequences, namely the efficient suppression of one type by the other.   

The paper is organized as follows: In the next section we discuss the evolution of the background system, and specify the different choices of free function ($V$, $C$ and $D$) forms and parameters used consistently throughout our analysis. In section 3 we will turn our attention to the evolution of cosmological perturbations in the presence of disformal and conformal couplings, and compute both matter and angular power spectra for various cases. We summarize our findings in section 4. All numerical work, including background simulations and both power spectra, is the output of a modified version of the publically available Boltzmann code CLASS \cite{Blas:2011rf}. Throughout the paper, we will emphasize coupling type discernabiliy; can we actually `observe' a purely disformal phenomenon?

\section{Background Cosmology}
\label{sec:background}

This section is split into several parts. Firstly, we write down the background equations. The background dynamics are then described in detail. The effective coupling to dark matter and the effective equation of state of the dark energy scalar field is discussed subsequently. 
	
\subsection{Equations of motion}
The background spacetime is the standard Friedmann--Robertson--Walker (FRW) metric solution to the Einstein equations for the metric $g_{\mu\nu}$ with flat spatial hypersurfaces: 
\be
\label{eq:FRW}
ds^2 = g_{\mu\nu}dx^{\mu}dx^{\nu} = a^2(\tau)[-d\tau^2 + \delta_{ij}dx^idx^j].
\ee 
Here $\tau$ is the conformal time and $a(\tau)$ is the scale factor. For the rest of the paper, dots denote derivatives with respect to $\tau$. 
Note that the disformal metric which dark matter particles "feel" is given by Eqns. \eqref{eq:dismetric} and \eqref{eq:FRW} as 
\be
\label{eq:JFline}
d\tilde{s}^2 = \tilde{g}_{\mu\nu}dx^{\mu}dx^{\nu} = Ca^2(\tau)[-\gamma^2d\tau^2 + \delta_{ij}dx^idx^j],
\ee 
where we define a disformal scalar $\gamma$ as 
\be
\label{eq:s}
\gamma:=\sqrt{1+\frac{D}{C}g^{\mu\nu}\phi,_{\mu}\phi,_{\nu}}.
\ee
The background value of the scalar field depends only on $\tau$. 
We assume that neutrinos are massless in our analysis, and hence the different sectors of the theory are specified by a massless relativistic component, $r$, and baryon component, $b$, a dark matter component $c$ and the scalar field $\phi$. The relativistic species as well as the baryons are assumed to be uncoupled from the scalar and hence the evolution of their energy densities is described by standard conservation equations:
\begin{subequations}
\label{eq:conservation}
\begin{align}
\label{eq:conservation:r}
\dot{\rho}_{\rm r} + 3{\mathcal H}(\rho_{\rm r} + P_{\rm r}) &= 0
\\
\label{eq:conservation:b}
\dot{\rho}_{\rm b} + 3{\mathcal H}\rho_{\rm b} &= 0
\end{align}
\end{subequations}
where $P_{\rm r} = \rho_{\rm r}/3$.
On the other hand, the scalar field obeys the Klein Gordon equation, and is now coupled to dark matter via a coupling function $Q$:
\be\label{eq:KGbackground}
\ddot \phi + 2 {\cal H}\dot\phi + a^2 \frac{dV}{d\phi} = a^2 Q~,
\ee
where the background form of $Q$ is given by 
\be\label{Qbackgroundcoupling}
Q = -\frac{a^{2}C'-2D(3{\cal H}\dot{\phi}+a^{2}V'+\frac{C'}{C}\dot{\phi}^{2})+D'\dot{\phi}^{2}}{2(a^{2}C+D(a^2\rho_c-\dot{\phi}^{2}))}\,\rho_c~.
\ee
The non-conservation of the dark energy--energy momentum tensor implies subsequent non-conservation of dark matter; energy loss from one species must be mirrored by energy gain in the other, and so for the cold dark matter species, c, we obtain 
\be
\dot\rho_{\rm c} + 3 {\cal H} \rho_{\rm c} = -Q \dot \phi~.
\label{eq:consbackground}
\ee
Finally, from Einstein's equations we present the Friedmann equation, which takes the standard form:  
\be\label{eq:Friedmann}
{\cal H}^2 = \frac{8\pi Ga^2}{3} \left( \rho_{\rm r} + \rho_{\rm b} + \rho_{\rm c} + \rho_{\rm de} \right)~,
\ee
where ${\cal H} = \dot a/a$ and $\rho_{\rm de} = \dot\phi^2/2a^2 + V(\phi)$. 


\subsection{Analysis of the dynamics}
The dark energy - dark matter interaction encoded in the equations above describes a peculiar scenario. For the dark energy field, the coupling contributes to an effective potential which depends in general on $\phi$ and $\dot{\phi}$. The dark matter then gains and loses energy as the geometry described by ${\tilde g_{\mu\nu}}$ is stretched and distorted; the conformal factor dilutes the dark matter over space-time by modifying the isotropic expansion it feels, $C^{1/2}a$, while the disformal factor distorts dark matter particles' light cones. The nature of this energy transfer process will depend crucially on how we specify our three free functions $V(\phi)$, $C(\phi)$ and $D(\phi)$. Suffice to say for now that we require our cosmology to be empirically plausible - observation tells us our dark energy field must roughly resemble a cosmological constant, and the scalar field must evolve very slowly to account for the accelerated expansion. To be specific, in this paper we study the following forms for $V$, $C$ and $D$: 
\begin{subequations}
\label{eq:free}
\begin{align}
\label{eq:free:V}
V &= M_V^4e^{\beta_V \phi}~, \\
\label{eq:free:C}
C &= C_0e^{\beta_C \phi},~ \\
\label{eq:free:D}
D &= M_D^{-4}e^{\beta_D \phi}~.
\end{align}
\end{subequations}
This choice represents the simplest extension to the standard coupled quintessence scenario, for which $C$ and $V$ are given as above and $D=0$. 

We can, without loss of generality, set $C_0=1$, as this parameter simply corresponds to a global redefinition of units. Such a choice does not affect the dynamics. The dark energy scale, $M_V$, is taken to be a fitting parameter that must be tuned such that our final time boundary conditions agree with measurement of the universe today, that is to say $\Omega_{\phi} = 0.68$, $\Omega_c = 0.27$ and $\Omega_b = 0.049$. Typically we find $M_V \sim {\mathcal H}_0 \sim {\rm meV}$. The conformal coupling is dimensionless, but the disformal factor introduces a new scale into the model. 
The case $M_D^{-1} \rightarrow 0$ corresponds to the standard coupled quintessence scenario, and the opposite limit where $M_D \rightarrow 0$, it turns out, is actually an uncoupled limit, regardless of the form of $C$. This unexpected feature is a consequence of a suppression effect to be clarified shortly. In between these limits, we find disformal effects leave an observable imprint on cosmological observables that is maximal if $M_D \approx M_V$.
We study a variety of models with different values for $M_D$, $\beta_C$ and $\beta_D$, and the parameter combinations we consider are summarized in table \ref{table:models}, where the meaning of the final column we will specify in the next section.

\begin{table}
\begin{center}
\ra{1.2}
    \begin{tabular}{@{} l l l l l l @{}}
    \toprule
    \# & Name       & $\beta_C$ & $\beta_D$~ & $M_D$    & $x$ behavior \\ \hline
    1  & Uncoupled ~& 0 ~       & 0         & $\infty$ & $x=1$ $\forall$ $\tau$ \\
    2  & Conformal ~& -0.2 ~    & 0  ~      & $\infty$~& decreasing with $\tau$    \\
    3  & Disformal ~& 0         & 0         & $M_V$  ~ & increasing with $\tau$    \\
    4  & Mixed     ~& -2        & 2 ~       & $M_V$    & single stationary minimum \\
\bottomrule
\end{tabular}
\caption{Description of the four models used throughout this paper. The model parameters are defined in Eqns. \eqref{eq:free}. The function $x$ is defined in Eqn. \eqref{eq:beta}, and `Name' corresponds to the model's label in plot legends. Note that in the fifth column, $M_D = \infty$ simply represents the limit for which the disformal coupling vanishes. For a more detailed discussion of the last column, we refer to Sec. 2.3.}
\label{table:models}
\end{center}
\end{table}

\begin{figure}
\begin{center}
\scalebox{0.6}{\includegraphics{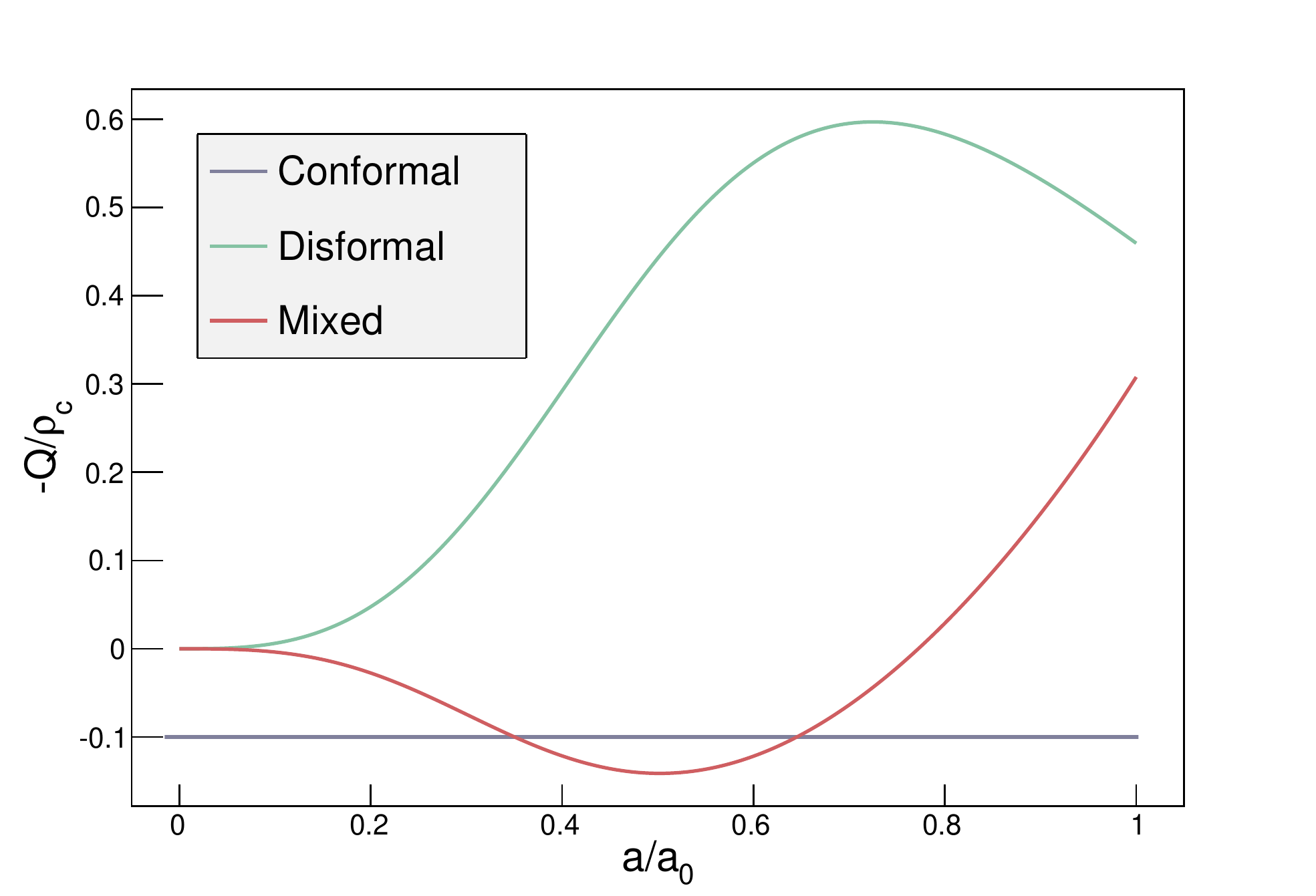}}
\caption{The scalar coupling function, $Q$, for background solutions of various models, plot against the scale factor $a$ where $a_0$ is the value of $a$ today. The free functions are defined in Eqns. \eqref{eq:free}, and $\beta_V = -2$ for all curves. The rest of the model parameters are given in table \ref{table:models}. }
\label{fig:Q_over_rho}
\end{center}
\end{figure}

\begin{figure}
\begin{center}
\scalebox{0.6}{\includegraphics{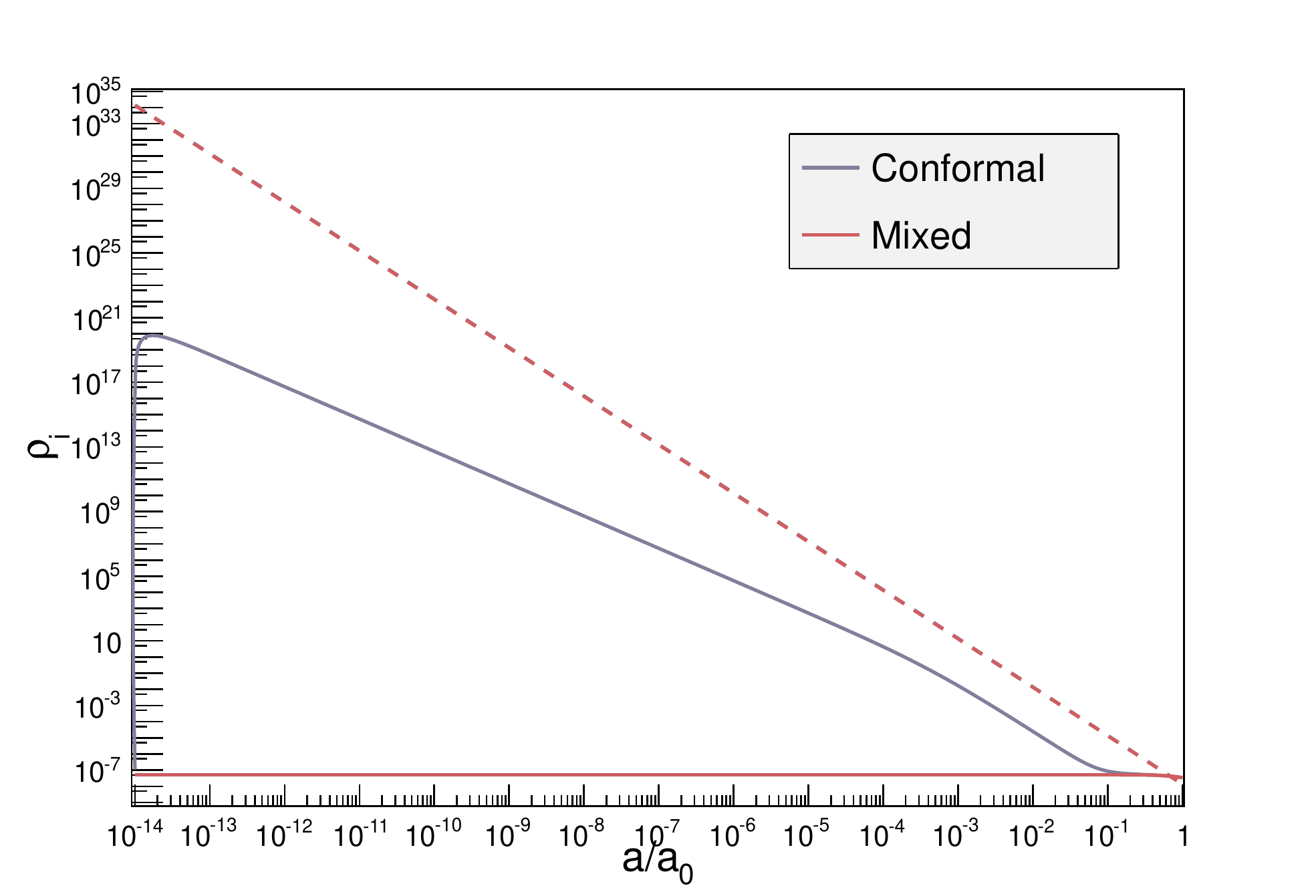}}
\caption{Evolution of energy densities of: dark matter (dashed lines) and dark energy (solid lines) for background solutions of various models, plot against the scale factor $a$ where $a_0$ is the value of $a$ today. The free functions are defined in Eqns. \eqref{eq:free}, and $\beta_V = -2$ for all curves. The rest of the model parameters are given in table \ref{table:models}. Here the dark matter curves for both models coincide. }
\label{fig:rhos}
\end{center}
\end{figure}

In Fig. \ref{fig:Q_over_rho}  we show the evolution of the coupling function, in which the most prominent feature is an early time suppression induced by the disformal factor. The coupling effectively `switches on' during some past epoch, quite late in the universe's evolution, and before this time it is in fact completely negligible. We see the effect this has on the evolution of the dark sector's energy densities in the next figure. Compared to the purely conformal case, the scalar field receives no great kick at early times when a disformal factor is included, and though $\beta_C$ is here a factor of 2 larger than current experimental upper bounds, the field mimics a cosmological constant throughout the majority of this universe's simulated lifetime.

Probing different free functions and parameters we find this suppression to be no lucky coincidence of the mixed model, but seems a general property of a disformal factor included in almost any cosmology. We can see why this must be the case by examining Eqn. \eqref{Qbackgroundcoupling}. A non-zero $D$ means the presence of a term proportional to $\rho_{\rm c}$ appears in the denominator. As long as the disformal scale $M_D$ is of the same order as the dark energy scale or less, this term will continue to make the coupling negligible until dark matter is roughly of that scale, i.e. today. We will show in later sections that the same screening effect holds too for linear perturbations, for the very same reason.

We have arrived at the first main result of the paper: in a cosmological setting, a disformal factor can suppress a conformal contribution at early times. The key point is that, for certain orders of magnitude of the disformal factor, the value of $Q$ and its linear perturbation, $\delta Q$ (whose exact form is given in section 3), are very much diminished for the majority of the universe's evolution - in the presence of disformal couplings, significant conformal ones are not necessarily in disagreement with cosmological observations. 

\subsection{The characteristic coupling function, $x$}
The expression of $Q$ is not simple to analyze, and collecting all our coupling effects under the obfuscated umbrella $Q$ has somewhat obscured the physics. It's form is necessary for finding numerical solutions, but for the analytics we can do better. To elucidate the effects conformal and disformal couplings have on cold dark matter, we will now define a new variable $x$ that greatly simplifies the analysis, and, as it turns out, the dark matter equation \eqref{eq:consbackground} will become easily solvable. In fact, this remains true for any species whose exact solution can be found in $\Lambda$CDM, for example photons. We relegate the details and general case to appendix \ref{sec:frame_transformations}, but for pressure-less dark matter we obtain 
\be\label{eq:rho_exact}
\rho_c = \rho_{0,c}a^{-3}x~,
\ee
where we have defined the quantity 
\be\label{eq:beta}
x:=\frac{C^{1/2}/\gamma}{C_0^{1/2}/\gamma_0}~.
\ee
Another useful quantity will be the derivative of $x$: 
\be\label{eq:beta_dot}
\frac{\dot{x}}{x} = - \frac{Q\dot{\phi}}{\rho_{\rm c}}~,
\ee
which we express in terms of a rate. Then, looking again at Eqn. \eqref{eq:consbackground}, we see now the evolution of $\rho_c$ as a competition of rates: that of the Hubble expansion rate, ${\cal H}$, and the rate of interaction with the scalar field, $\dot{x} / x$.

The positivity of $x$ follows naturally from the fact that both $C$ and $\gamma$ separately must be positive. This condition is defined by the metric \eqref{eq:JFline}; we must preserve causality, or suffer the consequences. Throughout the course of this paper, it will become clear that the whole system can be characterized by $x$ and its derivative - at least, in terms of observables - at both the background and perturbative level. We can already see this to be true at the zeroth order. As $x$ contains only background quantities however, it is quite remarkable that this remains true at first order.

With $x$ defined, it is now time for us to address the issue of our large free function space. What we are looking for in this study is general characteristics of conformal and disformally coupled dark matter, not idiosyncrasies corresponding to specific choices of the functional forms and parameters of $V$, $C$ and $D$. To make this step toward more comprehensive conclusions, we first notice that - as we have already stated - both $C$ and $D$ do not actually work independently, but affect the system jointly through $x$. We will then partition observationally distinct models based on the behavior of their respective $x$ function in conformal time.

As previously stated, we keep the models presented here realistic, with observables like the CMB anisotropies close to their measured values, and so we must work in the slow roll regime. This means that not only should $V$ be a relatively shallow function, but so too $C$, as it is also able to drive the field. The disformal factor, however, induces a damping in the field dynamics, and we find it can not push the field by itself, but rather hinders its movement. This damping will give us some more leeway in how shallow $C$ and $V$ can be. What this will ultimately mean is that we do not consider scaling solutions or attractors; this work is not aimed at solving the coincidence problem, rather, we find an alternative notion of naturalness is manifest here: the general inclusion of a disformal factor serves to push an arbitrary coupled cosmology toward one with a cosmological constant.

Given what has just been said, we suggest that for a qualitative first study, it will be enough to consider just four distinct models:
\begin{enumerate}
\item uncoupled quintessence, where $x=1$ $\forall$ $\tau$,
\item $x$ is a decreasing function  of time,
\item $x$ an increasing function of time,
\item $x$ has a single minimum.
\end{enumerate}
We can now comment on the final column of table \ref{table:models} mentioned earlier. In particular, it is not of vital importance that $V$ is of exponential form, a power law e.t.c. What does matter is whether $x$ is pushed upward, to larger values, or made to roll down to lower ones. So, only the direction of the slope of $C$ relative to $V$ affects the evolution of cold dark matter.

\begin{figure}
\begin{center}
\scalebox{0.6}{\includegraphics{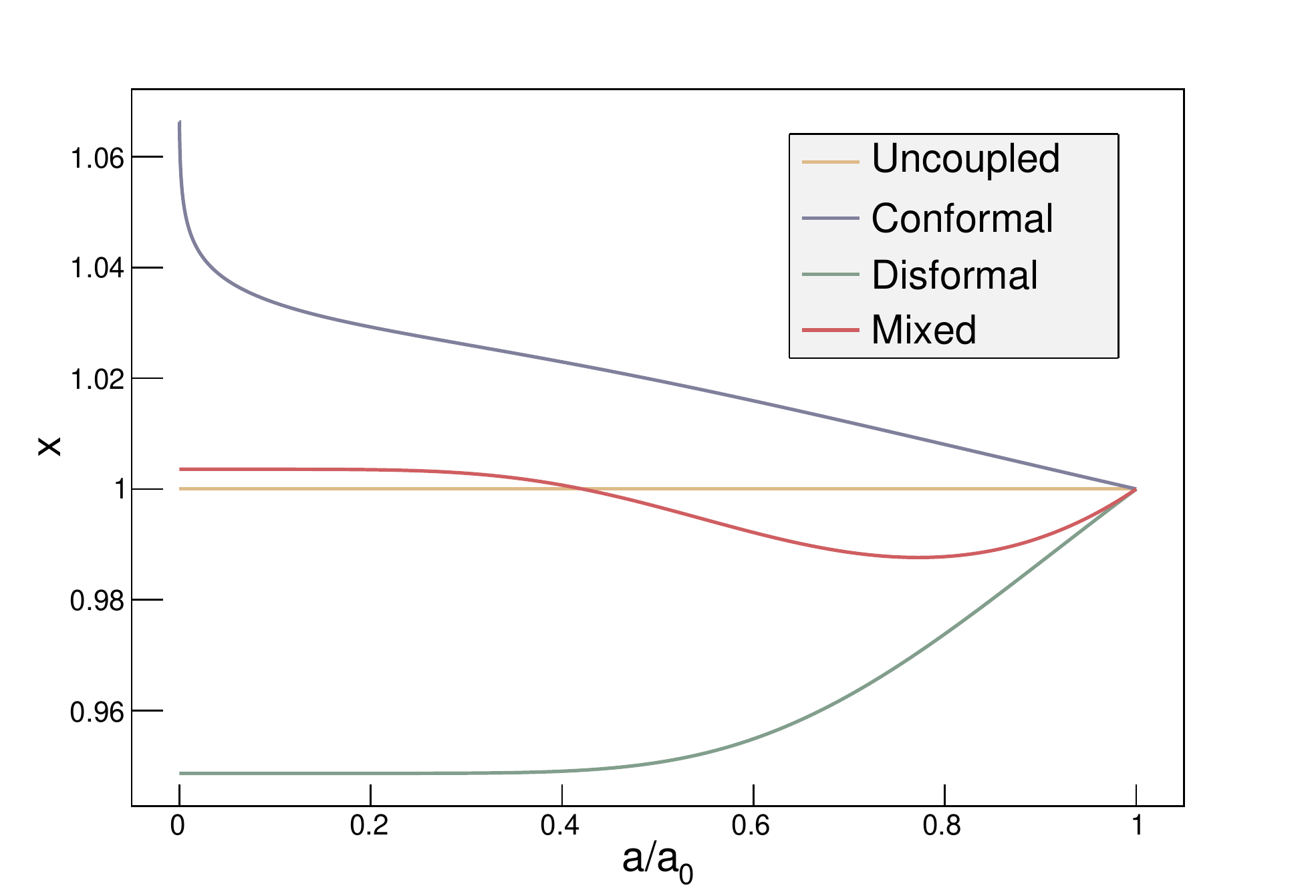}}
\caption{Evolution of the characteristic $x$ function for the four models described in table \ref{table:models}, with $\beta_V = -2$ for all curves. The model parameters are defined in Eqns. \eqref{eq:free}. }
\label{fig:beta}
\end{center}
\end{figure}

We show the $x$ behavior for the four models in figure \ref{fig:beta}. Note again the defining feature of the disformal term is that $\dot{x}$ - and hence the coupling - vanishes for early times, while for the conformal model it rapidly diverges. At early times then the distinction between conformal and disformal effects is strikingly clear, but at late times however, this is not so. We have used here a conformal only model to produce a decreasing $x$ function model, but we could have achieved this by other means.
For example, were we to pick a more complicated choice of $D$ function, and a different potential, we would get qualitatively the same late time behavior. At least at the background level, it is this late time $x$ behavior that is observable, as long as dark energy remains sub-dominant, and we will see in the next section why we have categorized our models based on this criteria.

\subsection{An effective equation of state}\label{sec:w_eff}
The dynamics of our gravitationally coupled system are in general quite complex. There is energy transfer between the elements of the dark sector that depends not only on the dark energy field, but also it's first and second derivatives. We anticipate that when interpreting data, cosmologists will use a much simpler parameterization. This assumed model is most often of a non-interacting dark sector, where dark matter is pressureless dust and dark energy some fluid with an open equation of state. Following \cite{Das:2005yj} we now reformulate our theory at the level of the zeroth order equations of motion to fit this neat picture, and define an effective, or apparent, dark energy equation of state $w_{\mathrm{eff}}$.

In the Friedman equation we first perform an effective splitting between the two dark components:
\be
{\mathcal H}^2 = \frac{8\pi Ga^2}{3} \left( \rho_{b,0} a^{-3} + \rho_{\rm{c},0}a^{-3} + \rho_{\rm de,eff} \right)
\ee
with 
\be
\rho_{\rm de,eff} = \rho_\phi - \rho_{\rm{c},0}a^{-3} + \rho_{\rm c}.
\ee
Then, taking the time derivative of $\rho_{\rm de,eff}$ and defining $w_{\rm eff}$ 
\begin{equation}
\dot\rho_{\rm de,eff} = -3{\mathcal H}\left( 1 + w_{\rm eff} \right)~,
\end{equation}
we find 
\be\label{eq:eoseff}
w_{\rm eff} = \frac{p_\phi}{\rho_{\rm de,eff}}~~~~~~{\rm with}~~~~~~p_\phi = X - V,
\ee
where $X=\dot\phi^2/2a^2$ is the kinetic energy of the field.
This quantity is the apparent equation of state of dark energy an observer would infer, if the assumption is made that the energy density of dark matter scales scales like $a^{-3}$. In our theory of course this is not the case, and $w_{\rm eff} \neq w_\phi$ in general.

What can we expect to observe in the behavior of this new effective system? An interesting first question to ask is: will we see phantom behavior? Using Eq. \eqref{eq:rho_exact}, it is simple for one to derive the following \emph{phantom condition}:
\be\label{eq:phan_con}
w_{\rm eff}<-1 ~~~~ \Leftrightarrow ~~~~ 2X<\rho_{\rm{c},0}a^{-3}(1-x)
\ee
where we recall that $x$ is normalized to unity today. Its clear that the evolution of $x$ will dictate whether or not we see the effective dark energy cross the phantom line, and this is directly related to the coupling of the underlying true model: if $x$ is an decreasing function phantom behavior is impossible, and in this scenario energy flow is from dark energy to dark matter. Conversely, energy flow in the opposing direction ($x$ is an increasing function) will propel the universe toward even greater expansion, as the relative contribution to the cosmic inventory from the vacuum energy will grow. Clearly, this system should not exhibit the standard instabilities expected from true phantom dark energy models, as it is simply a phenomenological re-parameterization.

\begin{figure}
\begin{center}
\scalebox{0.6}{\includegraphics{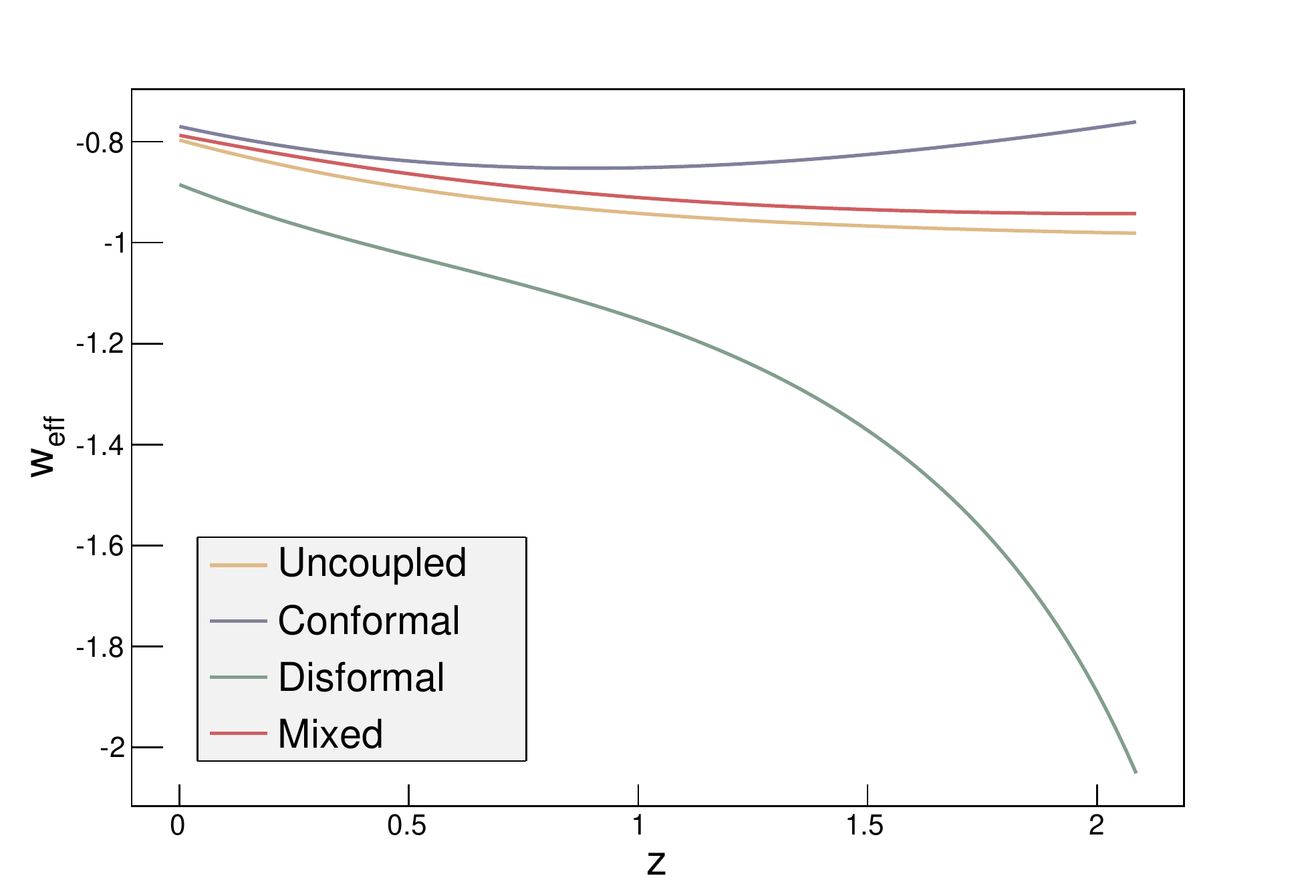}}
\caption{Evolution of the apparent equation of state, $w_{\rm eff}$, defined in Eqn \eqref{eq:eoseff} for the four models described in table \ref{table:models}. The model parameters are defined in Eqns.  \eqref{eq:free}. }
\label{fig:w_eff}
\end{center}
\end{figure}

For our four chosen models in table \ref{table:models}, we display the effective equation of state as discussed above in the redshift range accessible to the proposed Euclid satellite and the majority of current redshift galaxy surveys. While the disformal model here goes divergently phantom, for the others, the phantom line is never crossed. In the pure conformal case, the effective system tends further from $\Lambda$CDM, toward the boundary between acceleration and staticity. Das et al. \cite{Das:2005yj} however find a conformal model that replicates our disformal model's divergently phantom behavior, and so the take home message is emphatically \emph{not} that disformal couplings induce effective phantom behavior while conformal ones do not. Rather the point is that if $x$ is increasing with time, phantom behavior will likely ensue - a direct result of the phantom condition \eqref{eq:phan_con}.  

\begin{figure}
\begin{center}
\scalebox{0.6}{\includegraphics{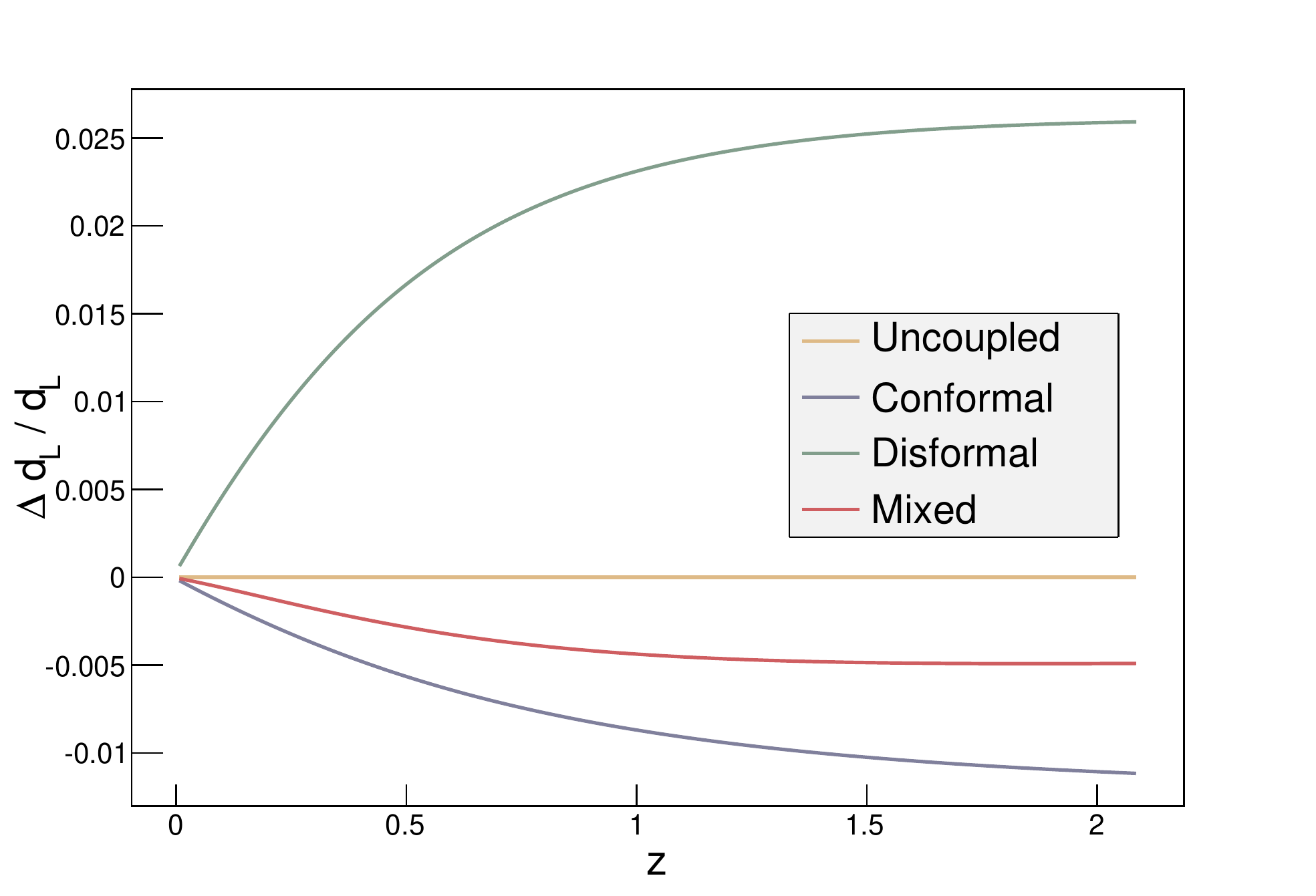}}
\caption{Evolution with redshift of the fractional difference in luminosity distance, Eqn. \eqref{eq:lum_dist}, between the four models described in table \ref{table:models} and the uncoupled case. The model parameters are defined in Eqns. \eqref{eq:free}. }
\label{fig:lum_dist}
\end{center}
\end{figure}

At the background level then, the coupling has a nice interpretation as a variable dark energy equation of state. This correspondence is best illustrated through its effect on luminosity distances, $d_{L}$. In figure \ref{fig:lum_dist} we show the luminosity distance difference ratio for our four models, defined as:
\be
\label{eq:lum_dist}
\frac{\Delta d_{\rm L}}{d_{\rm L}} \bigg |_{\rm i} := \frac{d_{\rm L,i}-d_{\rm L, uncoupled}}{d_{\rm L, uncoupled}}.
\ee
As our intuition suggests, energy flow into the scalar field accelerates expansion, causing observed objects such as supernova at a fixed redshift to appear further from us than for the uncoupled case.

To conclude this section, we have shown that conformal and disformal effects can not always be distinguished when only dark matter is coupled. Whether they can or not typically depends on the epoch in question: at early times the distinction is clear, as disformal contributions in general suppress conformal ones; at late times the two act together through the $x$ function, and whether energy flow is into or out of dark matter tells us nothing about the underlying behavior of $C$ and $D$, nor will any observed phantom behavior. What defines early and late times in this context is the new scale introduced by the disformal factor, $M_D$. When the dark matter energy density becomes comparable to that scale, the coupling switches on and begins to influence the field.

The reason why the two types of coupling become indistinguishable at late times is ultimately because dark matter is pressureless. What sets disformal factors apart, we have seen, is that they warp light cones and shift causal structure, but this has little effect when the particle constituents of the coupled fluid is cold dark matter.

\section{Evolution of cosmological perturbations}
\label{sec:perturbations}
We now turn our attention to the evolution of cosmological perturbations in our theory. We begin by first writing down the perturbation equations and subsequently discuss predictions for cosmological observables, such as the CMB anisotropies and matter power spectrum. Along the way we will try to be categorical about the various effects induced by the couplings for the different models; will there be anything about these spectra that is characteristically disformal?

To be concrete, we will be working in the Newtonian gauge. To avoid confusion, we reserve $\delta$ to denote matter density contrast: $\delta := \frac{\delta \rho}{\rho}$, $\delta P$ the pressure perturbation and $\delta\phi$ is the perturbation of the scalar field. We denote by $\hat\delta$ a general perturbation operator. The perturbed Einstein frame line element in the chosen gauge is:
\be
\label{eq:pert_metric}
ds^2 = a^2(\tau)\left[ -(1+2\Psi)d\tau^2 + (1-2\Phi)\delta_{ij}dx^idx^j \right]
\ee
which means that dark matter particles follow geodesics described by the following perturbed space-time:
\be
\label{eq:pert_metric_dis}
d\tilde{s}^2 = Ca^2(\tau)[-(1+2A)\gamma^2d\tau^2
	+ 2(\partial_iB)\gamma d\tau dx^i
	+ (1-2E)\delta_{ij}dx^idx^j]
\ee
where $A$, $B$, and $E$ are functions of the dark energy field background and perturbation values. Their exact forms are: 
\begin{subequations}
\label{eq:pert_line}
\begin{align}
\label{eq:pert_line:A}
A &= \Psi + \frac{\hat{\delta} C}{2C} + \frac{\hat{\delta} \gamma}{\gamma} \\
\label{eq:pert_line:B}
B &= \left(\frac{1}{\gamma}-\gamma\right)\frac{{\delta} \phi}{\phi'} \\
\label{eq:pert_line:E}
E &= \Phi - \frac{\hat{\delta} C}{2C}
\end{align}
\end{subequations}
Its now clear that only disformal factors induce off diagonal perturbations in the metric which, we will see, affect the velocity field perturbations.

For the visible sector we again neglect that neutrinos have mass. The perturbation equations of relativistic, $r$, and baryonic, $b$, matter follow from the energy--momentum 
conservation equations and are given by 
\begin{subequations}
\begin{align}
\dot{\delta}_r &= -(1+w_r)(\theta_r -3\dot{\Phi}) - 3{\mathcal H}\left(\frac{\delta P_r}{\delta \rho_r}-w_r\right)\delta_r 
\\
\dot{\theta}_r &= - {\mathcal H}(1-3w_r)\theta_r - \frac{\dot{w}_r}{1+w_r}\theta_r + k^2\Psi 
 + \frac{\delta P_r/\delta\rho_r}{1+w_r}k^{2}\delta_r 
\\
\nonumber\\
\dot{\delta}_b &= - \theta_b + 3\dot{\Phi} 
\\
\dot{\theta}_b &= -{\mathcal H}\theta_b + k^2\Phi.
\end{align}
\end{subequations}
Perturbations in the dark energy field, $\delta \phi$, evolve according to the perturbed Klein Gordon equation
\begin{equation}
\label{eq:KGperturbed}
\delta\ddot{\phi}+2{\cal H}\delta\dot{\phi}+(k^{2}+a^{2}V'')\delta\phi
= \dot{\phi}(\dot{\Psi} + 3\dot{\Phi})
- 2a^{2}(V'-Q)\Psi
+ a^{2}\delta Q
\end{equation}
and perturbation of $Q$ is given by the cumbersome expression \cite{vandeBruck:2012vq}
\be
\label{eq:Qperturb}
\delta Q = -\frac{\rho_c}{a^{2}C+D(a^{2}\rho_c-\dot{\phi}^2)}[{\cal B}_1\delta+{\cal B}_2\dot{\Phi}+{\cal B}_3\Psi+{\cal B}_4\delta\dot{\phi}+{\cal B}_5\delta\phi],
\ee
where
\begin{subequations}
\begin{align}
{\cal B}_1 &= \frac{a^{2}C'}{2} - 3D{\cal H}\dot{\phi} - Da^{2}(V'-Q) 
           - D\dot{\phi}^2\left(\frac{C'}{C}-\frac{D'}{2D}\right), \\
{\cal B}_2 &= 3D\dot{\phi}, \\
{\cal B}_3 &= 6D{\cal H}\dot{\phi}+2D\dot{\phi}^2\left(\frac{C'}{C}-\frac{D'}{2D}+\frac{Q}{\rho_c}\right), \\
{\cal B}_4 &= -3D{\cal H}-2D\dot{\phi}\left(\frac{C'}{C}-\frac{D'}{2D}+\frac{Q}{\rho_c}\right), \\
{\cal B}_5 &= \frac{a^{2}C''}{2}-Dk^{2}-Da^{2}V'' 
           - D'a^{2}V'-3D'{\cal H}\dot{\phi} 
           - D\dot{\phi}^2\left(\frac{C''}{C}-\left(\frac{C'}{C}\right)^{2}+\frac{C'D'}{CD}-\frac{D''}{2D}\right) \nonumber \\
           &+ (a^{2}C'+D'a^2\rho_c-D'\dot{\phi}^{2})\frac{Q}{\rho_c}~.
\end{align}
\end{subequations}
The non-conservation of dark matter induces, at the perturbative level, factors of $Q$ and $\delta Q$, in its conservation equations, which become
\begin{subequations}
\begin{align}
\dot{\delta}_c &= 
- \theta_c 
+ 3\dot{\Phi} 
+ \frac{Q}{\rho_c}\dot{\phi}\,\delta_c
- \frac{\dot{\phi}}{\rho_c}\delta Q
- \frac{Q}{\rho_c}\dot{\delta\phi} ,
\label{one}
\\
\dot{\theta}_c &= 
-{\cal H}\theta_c 
+ k^2\Psi 
+\frac{Q}{\rho_c}\dot{\phi}\theta_c
-\frac{Q}{\rho_c}k^2\delta\phi.
\label{two}
\end{align}
\end{subequations}

An important point has to be clarified before we continue: are we justified in setting the Einstein frame pressure and shear of the coupled dark matter to zero? After all, its been shown previously that disformal couplings can transform the equation of state of a coupled species to one dependent on the field, $\phi$ (see \cite{vandeBruck:2013yxa}, \cite{Minamitsuji:2014waa} and appendix \ref{sec:frame_transformations}). In the appendix we present the full set of frame transformations between Jordan and Einstein frame matter variables (or equivalently, between uncoupled and coupled variables respectively)\footnote{By Jordan frame, we mean the frame in which the action \eqref{eq:action} is rewritten fully in terms of ${\tilde g}_{\mu\nu}$, whereas the Einstein frame is specified by \eqref{eq:action}.}. In particular, for a general coupled species:

\begin{subequations}
\begin{align}
\tilde{\delta P} &= \frac{1}{C^2\gamma}\left[\delta P - \left(\frac{2\hat{\delta}C}{C} + \frac{\hat{\delta}\gamma}{\gamma}\right) P\right]~,
\\
\tilde{\Pi}^i_j &= \Pi^i_j~.
\end{align}
\end{subequations}
So, as dark matter has vanishing pressure in the frame for which it is uncoupled, we may indeed set $\delta P_c, ~ \Pi^{ij}_c = 0$.

The perturbed Einstein equations in our theory take the form as in standard $\Lambda$CDM. We don't quote them all here, but instead present just those important for our analysis of structure growth.
In particular, we will make use of the $00-$component of the Einstein equations
\be
\label{eq:phi_mt}
k^2\Phi + 3{\mathcal H}\left( \dot{\Phi} + {\mathcal H}\Psi \right)
= - 4\pi G a^2 \delta \rho
\ee  
and the $i\neq j-$component, which leads to 
\be\label{eq:slip}
\Phi - \Psi = 0 ~~~~\forall~~ C,D,
\ee
as we ignore anisotropic stress (we remind the reader that we ignore neutrino masses in our analysis). Our theory thus predicts no gravitational slip $(\eta := \Phi/\Psi = 1)$, independent of the coupling.  

\begin{figure}
\begin{center}
\scalebox{0.6}{\includegraphics{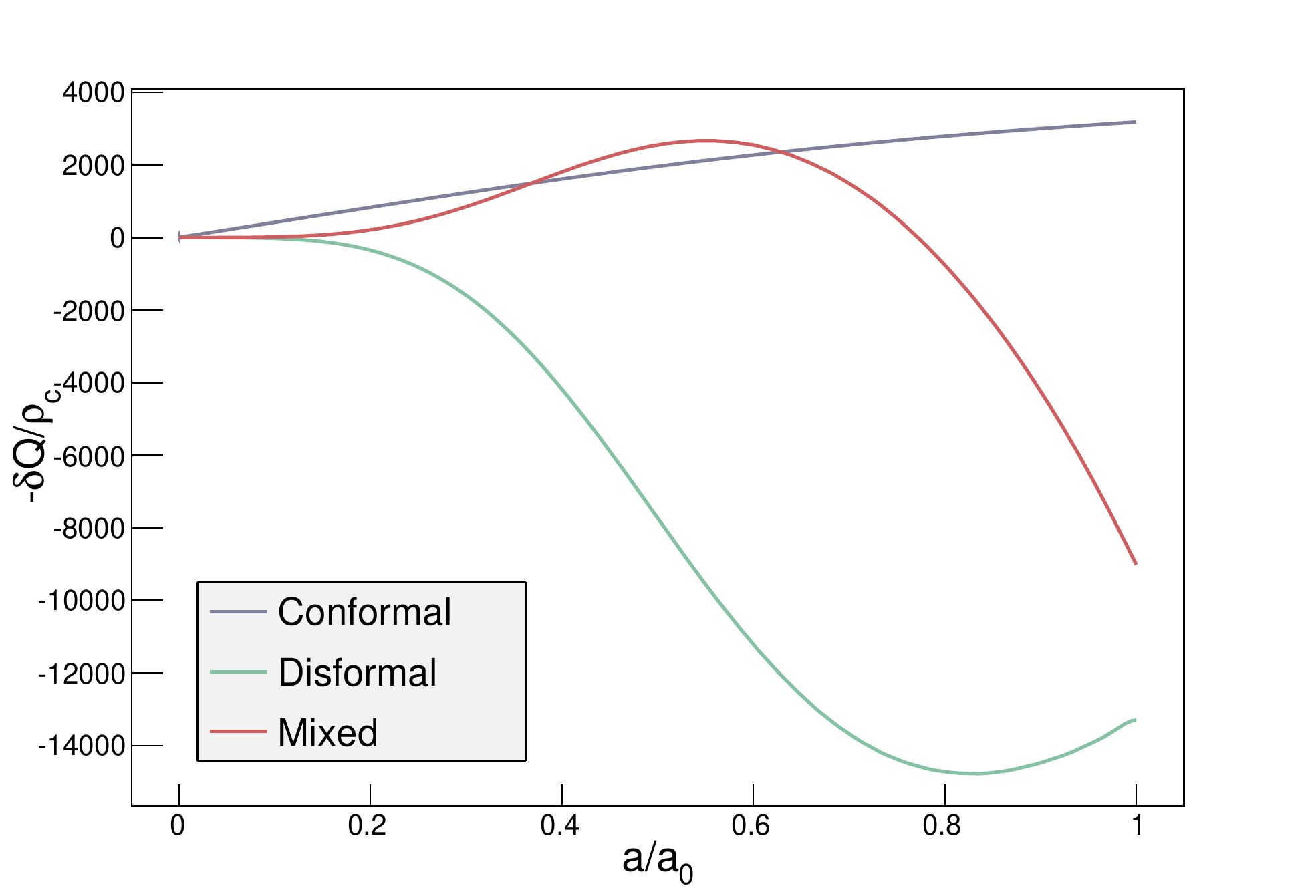}}
\caption{Dependence on the scale factor of the linear perturbation in the coupling function, $\delta Q$, given by Eqn. \eqref{eq:Qperturb}. The three models shown are described in table \ref{table:models}, with equations \eqref{eq:free}. }
\label{fig:delta_Q_over_rho}
\end{center}
\end{figure}

\begin{figure}
\begin{center}
\scalebox{0.6}{\includegraphics{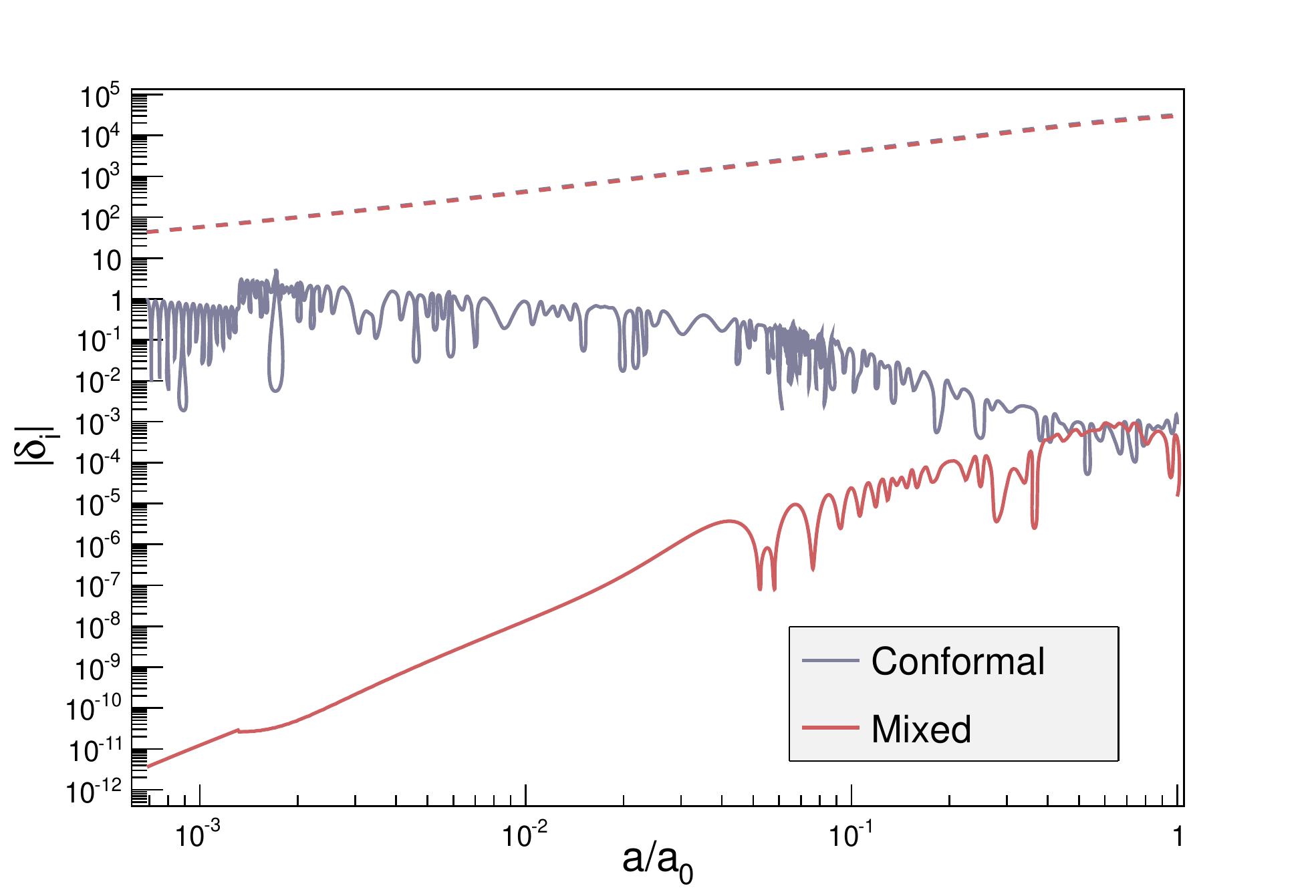}}
\caption{Evolution with the scale factor of density contrast absolute values in: the dark matter energy density (dashed lines), and dark energy density (solid lines). These curves correspond to modes with wavenumber $k=0.3Mpc^{-1}$. The models shown are described in table \ref{table:models}, with Eqns. \eqref{eq:free}.}
\label{fig:deltas}
\end{center}
\end{figure}

With the perturbed equations written down, we first ask whether the disformal factor suppresses a conformal one at the level of linear perturbations. Before getting to the figures, we can already guess that it will be the case from Eqn. \eqref{eq:Qperturb}; looking at the denominator we see the same $Da^2\rho_c$ term that was responsible for the suppression at zeroth order. Still, its important to be concrete, and so we demonstrate this suppression in Fig. \ref{fig:delta_Q_over_rho} and Fig. \ref{fig:deltas}.

\begin{figure}
\begin{center}
\scalebox{0.6}{\includegraphics{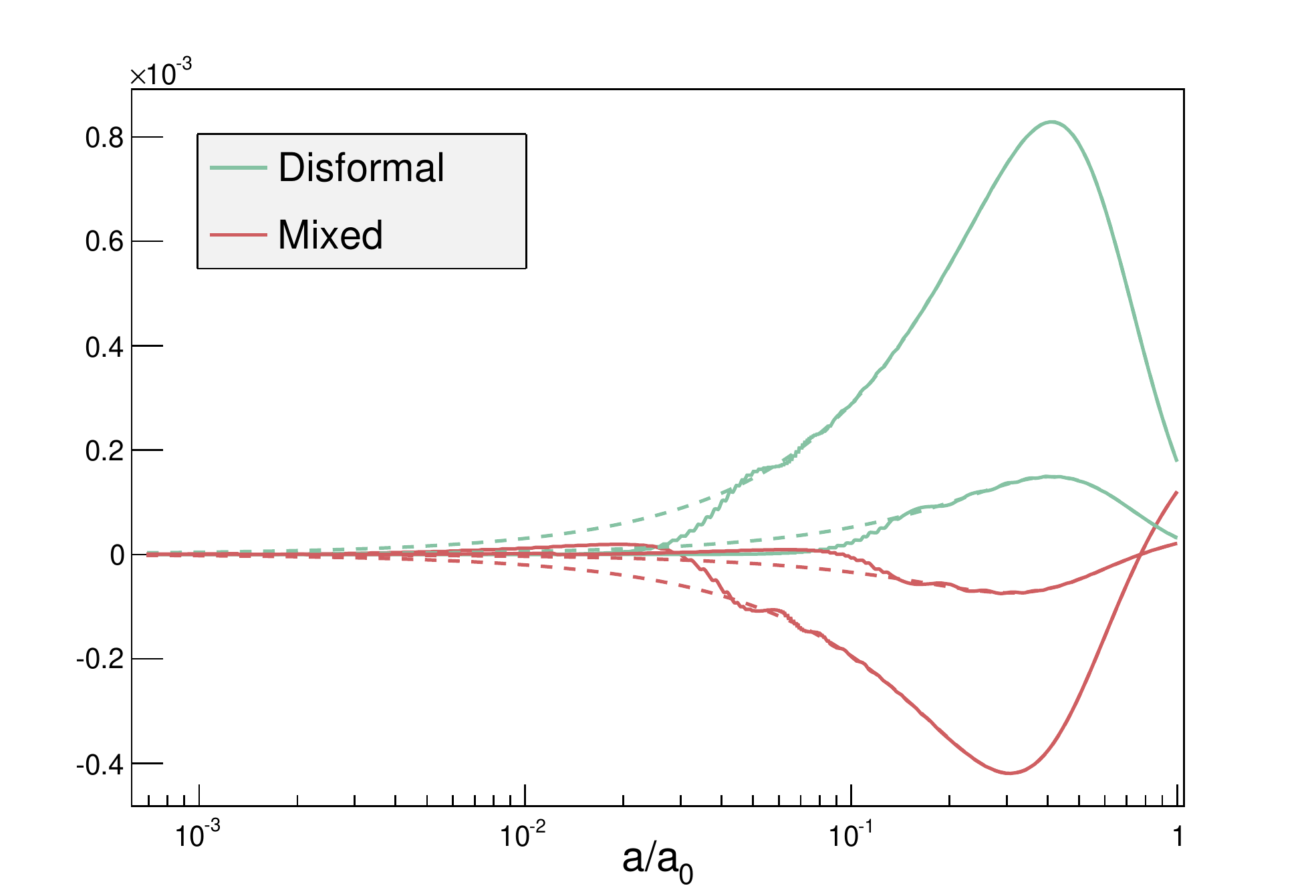}}
\caption{We show here, for models only including a disformal term, how the system approaches Eqn. \eqref{eq:delta_Q}. Dashed lines: $Q\delta_c$, where Q is defined by Eqns. \eqref{Qbackgroundcoupling}, and $\delta_c$ is the dark matter density contrast. Solid lines: $\delta Q$, given by \eqref{eq:Qperturb}. The models shown are described in table \ref{table:models}, along with Eqns. \eqref{eq:free}.}
\label{fig:delta_Q}
\end{center}
\end{figure}

When we compare figures \ref{fig:Q_over_rho} and \ref{fig:delta_Q_over_rho} a similarity immediately leaps out: one plot appears to resemble the negative of the other. Additionally, the curves for $\delta Q$ appear to rise and fall gradually, over background timescales, rather than, for example, the fast oscillations in the field perturbations (figure \ref{fig:deltas}). This simplicity may come as a surprise when juxtaposed with the arduous complexity of the $\delta Q$ equation from which the curves have sprung. In fact there is a strong relationship between the background and perturbed coupling function $Q$, and the damped oscillatory behavior of $\delta Q$ at $a/a_0 \sim 2\times 10^{-2}$ (figure \ref{fig:delta_Q}) betrays a important aspect of the coupling - the system at late times is drawn to a solution where \cite{Zumalacarregui:2012us} 
\be
\label{eq:delta_Q}
\delta Q \simeq Q\delta_c,
\ee
which we identify with the limit in which all dark energy perturbations in $\delta Q$ vanish.
\begin{figure}
\begin{center}
\scalebox{0.6}{\includegraphics{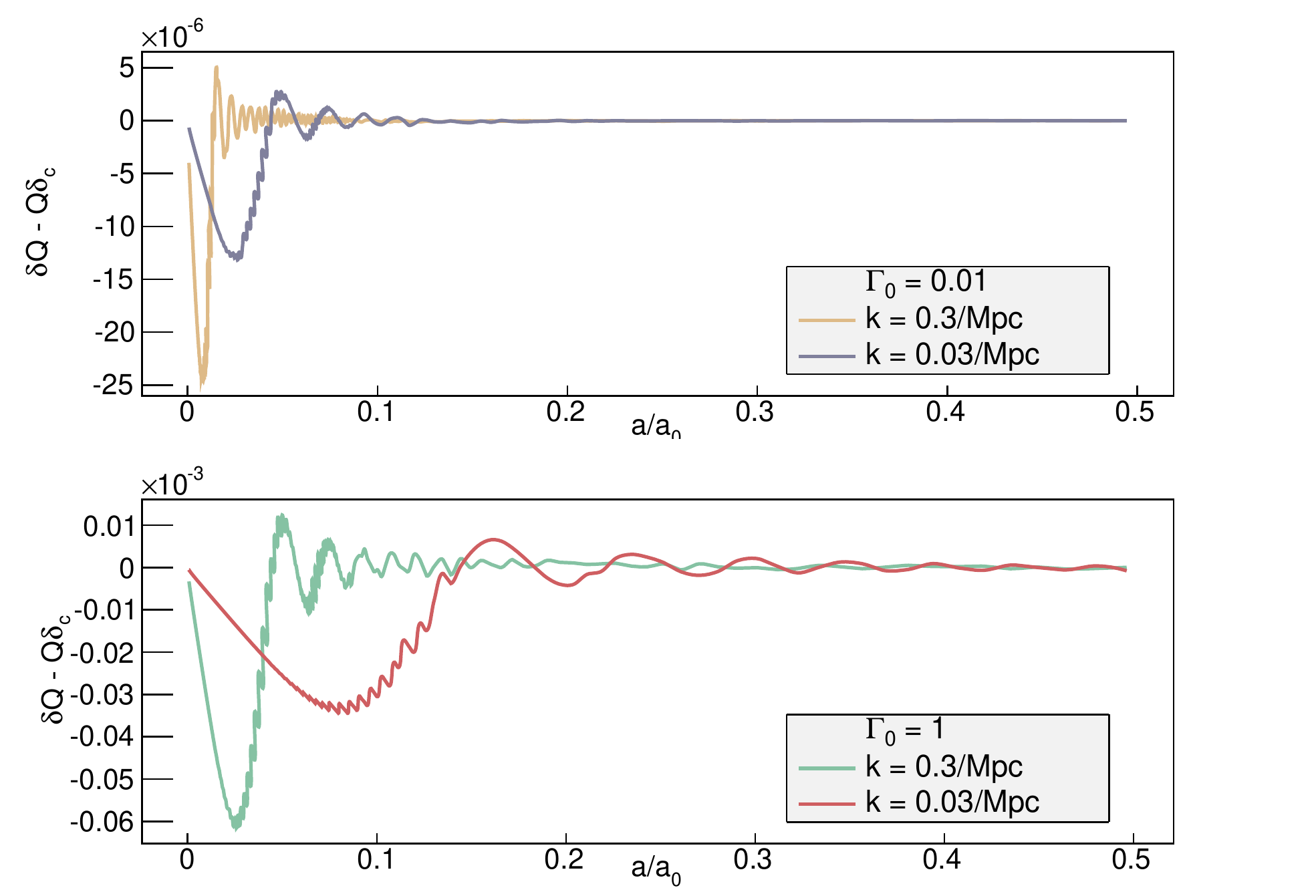}}
\caption{Oscillations induced by a disformal term in the perturbed coupling function \eqref{eq:Qperturb} for models where $\beta_V = -2$, $\beta_C=0$, $\beta_D=0$ and $M_D = M_V/\Gamma_0^{1/4}$. Terms in $\delta Q$ proportional to dark matter perturbations only, $Q\delta_c$, have been subtracted to isolate the dark energy perturbation fluctuations.}
\label{fig:deltaQ_minus}
\end{center}
\end{figure}
The approximate equivalence, Eqn. \eqref{eq:delta_Q}, does not hold at very early times but becomes progressively more accurate later on. In Fig. \ref{fig:deltaQ_minus} we plot the oscillating part of $\delta Q$ for an exclusively disformal model. We see clearly that both the oscillation and damping time scales depend on the mode's wavenumber, $k$, which we expect, and the new scale introduced by the disformal factor, $M_D$. In the conformal limit, characterized by $M_D \rightarrow \infty$, we find in general they disappear as their period tends also to $\infty$: a conformal factor in general introduces no scale, so a purely conformally coupled theory does not exhibit these oscillations. 

To briefly summarize by way of analogy, at the background level the new disformal scale determined an epoch in which the coupling is effectively `turned on'; at the perturbation level, $M_D$ now sets a damping and oscillating timescale for perturbations in the dark energy field. The introduction of this scale, we can now see, is primarily what sets apart the effects of conformal and disformal factors on the scalar field dynamics at the zeroth and linear order levels of the theory.

\subsection{Growth of large scale structure}
We have seen some evidence to support that, at least in principle, conformal and disformal effects \emph{can} be separated. The new important feature is of course the new scale, $M_D$. We now turn to the pressing question of observables: does the new scale also leave recognizable imprints on the formation and subsequent growth of structure? 

The key quantity we are interested in is the growth factor, $f(z)$, which is a convenient parameterization of linear growth. In the literature, the growth rate is defined as:
\be
\label{eq:f}
f(z) := \frac{{\rm d}\ln \delta_{\rm m}}{{\rm d}\ln a}=\frac{\dot \delta_{\rm m}}{\mathcal{H}\delta_{\rm m}},
\ee
where we notice the definition is with respect to all matter, $\delta_m := \delta\rho_m/\rho_m$ for $\rho_m = \rho_c + \rho_b$, not just dark matter. As our theory only couples dark matter to the scalar, we are now faced with an interesting question: how will the composite fluid of dark matter and baryons cluster into structure, if both species feel different (effective) gravitational forces in general? We would like to calculate the growth equation for not just cold dark matter, but for all matter. It is this total matter quantity that influences the gravitational potentials, which lens distant galaxy and CMB light. As a first step then, we now define a `baryon bias' parameter $b$, as: 
\be
\label{eq:bias}
\delta_b = b\delta_m
\ee
where $b$ will in general depend on time (and possibly scale, but since the scalar field is nearly massless, we find that $b$ in our model does not depend on the wave number $k$; the situation would be different for different potentials which are not of quintessence form). 

\begin{figure}
\begin{center}
\scalebox{0.6}{\includegraphics{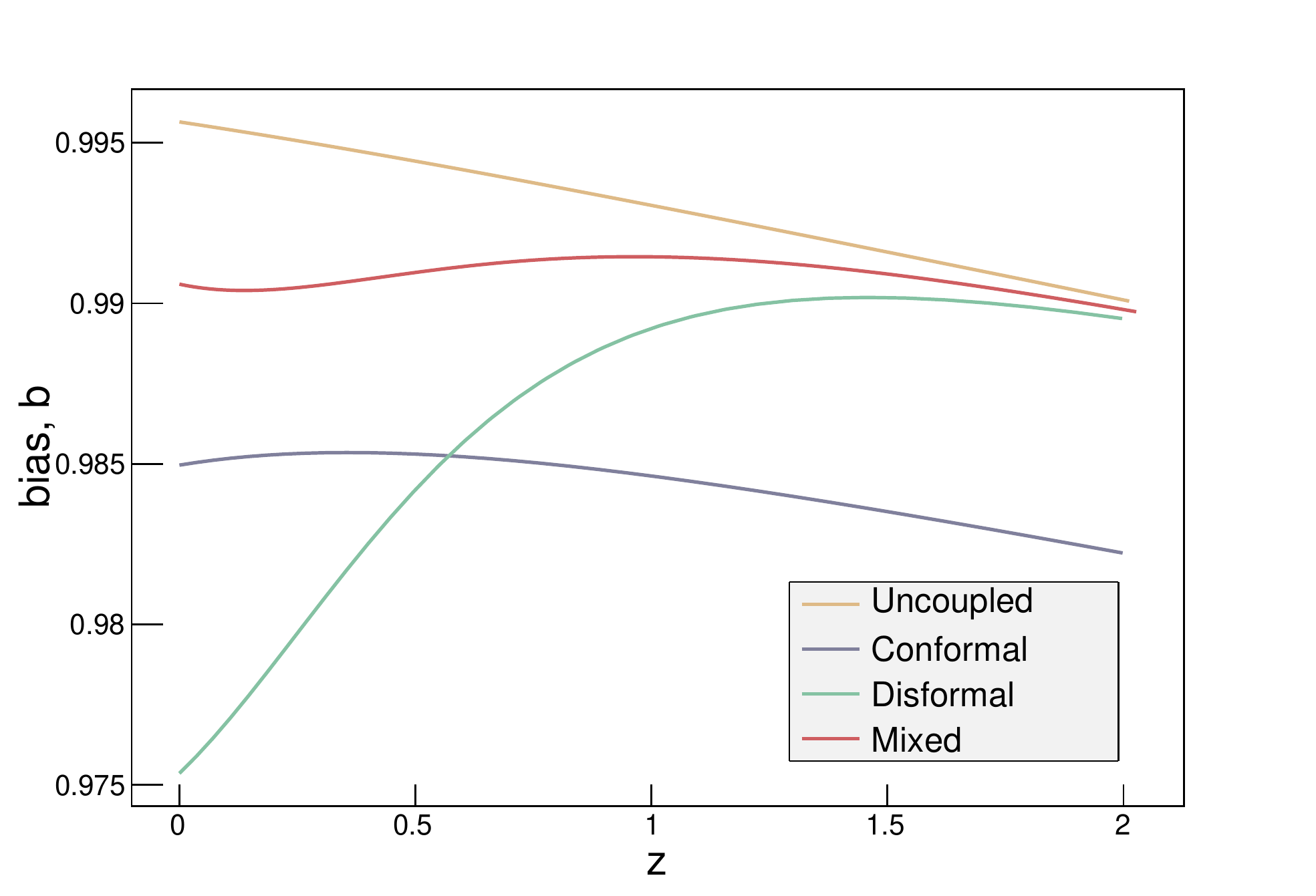}}
\caption{How the late time bias, as defined in Eqn. \eqref{eq:bias}, varies with redshift, z, for the models described in table \ref{table:models}. The free functions that define the parameters displayed in the table are given by Eqns. \eqref{eq:free}.}
\label{fig:bias}
\end{center}
\end{figure}

If we plot the late time evolution of this bias (figure \ref{fig:bias}) we see as we might expect that dark matter and baryons cluster at different rates, reflecting the underlying variations in each species' experience of gravity. What, primarily, we glean from the plots is that to reasonable accuracy, $\delta_b \simeq \delta_m$; from this we directly infer also that $\delta_c \simeq \delta_m$. We will return to address the validity of this assumption very soon, but for now, using this, along with sub-horizon approximations and Eqn. \eqref{eq:delta_Q} let us derive the linearized growth equation for matter when cold dark matter is gravitationally coupled to dark energy.

Let us take for granted momentarily that $\delta_c \simeq \delta_m$. By the additivity of the stress energy tensors and baryon conservation, the evolution of all-matter perturbations (dark matter + baryons) can be derived from:
\be
\hat{\delta}\left( \nabla_{\mu}T^{\mu\nu}_m \right) = \hat{\delta}\left( \nabla_{\mu}T^{\mu\nu}_b + \nabla_{\mu}T^{\mu\nu}_c \right) = \hat{\delta}\left( Q\phi^{\nu} \right)
\ee
which gives the growth equation:
\begin{subequations}
\be
\label{eq:growth_equation}
\ddot{\delta}_m + \mathcal{H}_{\rm eff}\dot{\delta}_m 
	= 4\pi G_{\rm eff}a^2\rho_m\delta_m
\ee
where we have defined:
\begin{align}
{\mathcal H}_{\rm eff} &:=\mathcal{H} + \frac{\dot{x}}{x}\frac{\rho_c}{\rho_m}\, \\
G_{\rm eff} &:= G + \frac{1}{4\pi \dot{\phi}^2}\left(\frac{\dot{x}}{x}\frac{\rho_c}{\rho_m} \right)^2\ = G + \frac{1}{4\pi}\frac{Q^2}{\rho_m^2},
\end{align}
\end{subequations}
which is valid for $k \gg {\mathcal H}$. We note that while the error $|\frac{\delta_m - \delta_b}{\delta_m}|$ is of the order $\sim1-3\%$ for the models considered, the error this induces in the growth equation \eqref{eq:growth_equation} turns out to be only ever as large as $\sim0.1\%$ in general, and usually substantially smaller. The error propagates through the derivation in a favorable way, affording us valuable comparison between the true evolution of 
$\delta_{\rm m}$ and our simplified growth equation to an accuracy sufficient for this study. We stress that, of course, evolution of cosmological perturbations is inextricably linked to evolution of the background.

\begin{figure}
\begin{center}
\scalebox{0.6}{\includegraphics{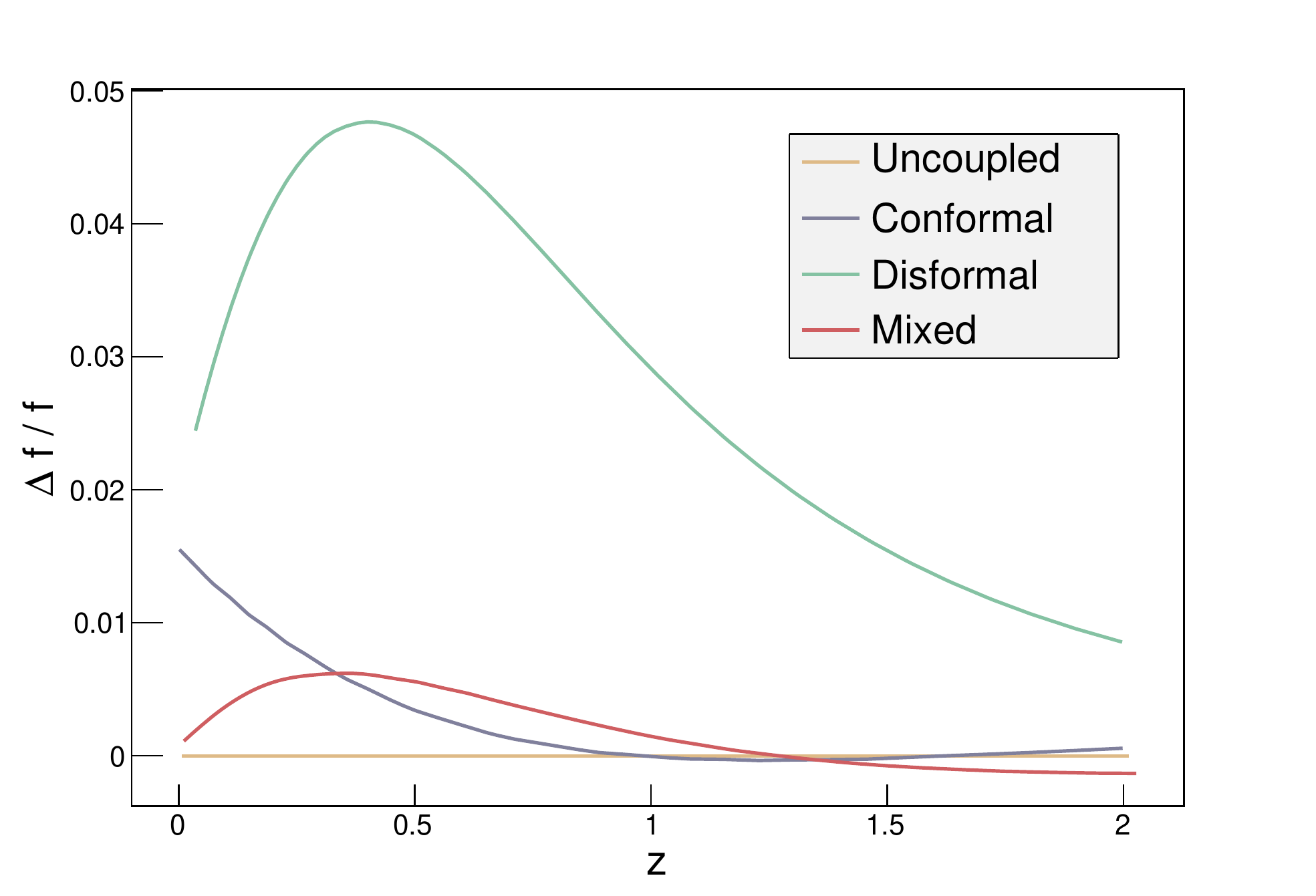}}
\caption{The fractional difference in the growth rate $f(z)$, Eqn. \eqref{eq:f}, for the models of table \ref{table:models}, against redshift, z. The difference is with respect to the uncoupled case, i.e. $\Delta f / f:= (f_i-f_{\rm uncoupled}) / f_{\rm uncoupled}$ where i runs over the four models.}
\label{fig:growth_vs_GR}
\end{center}
\end{figure}

We are now in a position to examine the growth rate, \eqref{eq:f} for our various models. First to note is that both $\mathcal{H}_{\rm eff}$ and $G_{\rm eff}$ contain only background quantities that have no $k$ dependence - any departure from general relativity here will be scale invariant. This is due to our choice of potential for the scalar field, which is of quintessence form and the field is nearly massless. With this choice, our theory predicts that, like $\Lambda$CDM, measuring the growth at different length scales (within the quasi-static regime of course) will not lay bare the novel features of the coupling presented here. If the oscillations depicted in Fig. \ref{fig:deltaQ_minus} were perhaps to have survived till today, this would change the story: a major observational test to distinguish disformal couplings would then be to see how these fluctuations depend on scale, $k$, and thus probe the value of $M_D$ itself. In general though this is not the case; the severe damping present in all models we have considered show that sustaining these oscillations long enough to observe them is difficult to achieve in practice, and so highly unlikely in reality. What looked in previous sections like a tool to measure disformal couplings turned out to be just an fleeting fluctuation, completely intractable empirically.

Looking at the curve for the purely conformal case in Fig. \ref{fig:growth_vs_GR}, it seems that, because both $\mathcal{H}$ and $\mathcal{H}_{\rm eff}$ are suppressed by the coupling (the background expansion rate is slowed as energy is transferred from dark energy to dark matter, exemplified by the effective equation of state for this model) growth is enhanced by the coupling. For the disformal only model, cosmic expansion \emph{and} the extra friction felt by $\delta_m$ ($\mathcal{H}_{\rm eff}$) is enhanced but we see that the growth rate is largest for the purely disformal model.  

\begin{figure}
\begin{center}
\scalebox{0.6}{\includegraphics{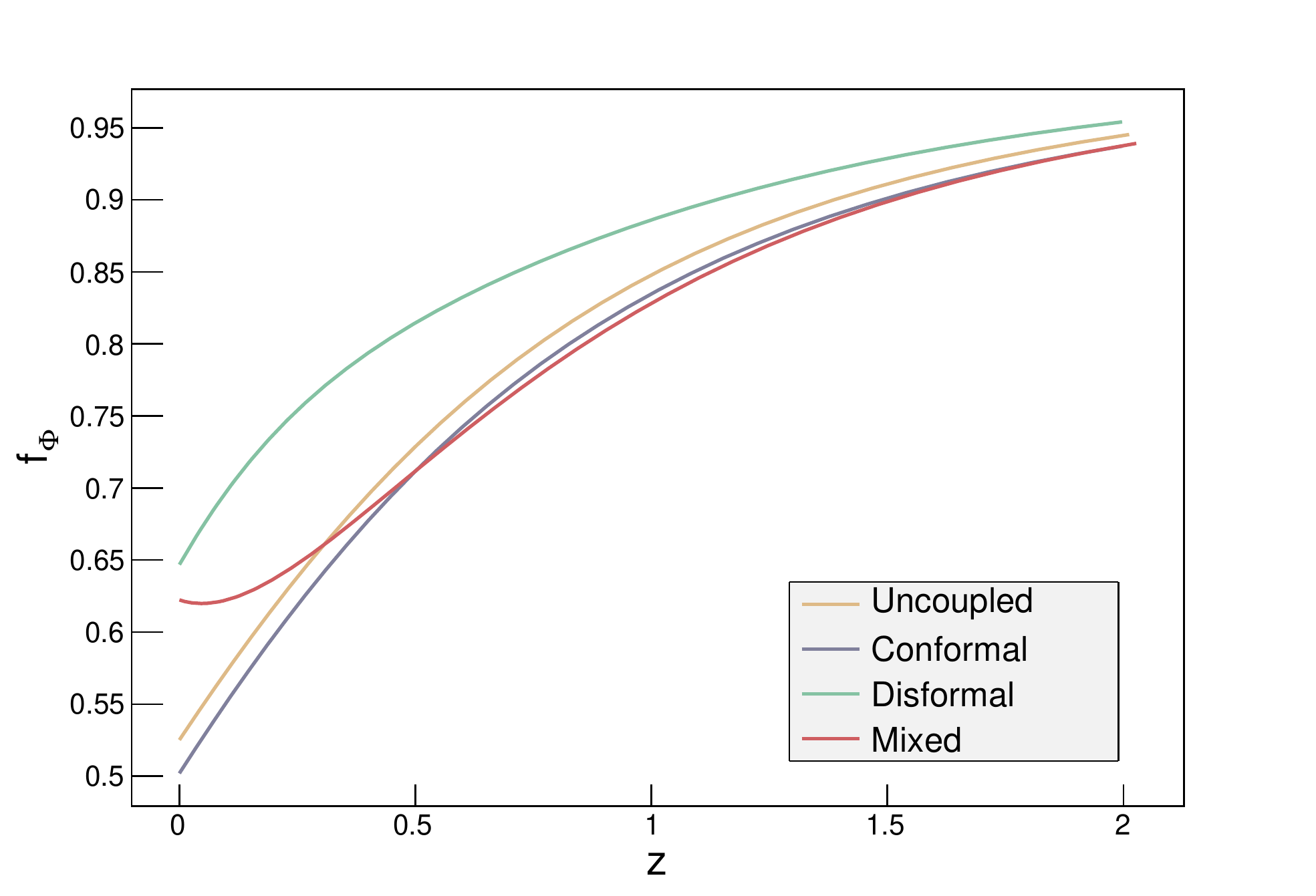}}
\caption{Evolution of the growth rate $f_\Phi$, as defined in Eqn. \eqref{{eq:f_eff}},  as a function of redshift, z, for the models of table \ref{table:models}.}
\label{fig:f_eff}
\end{center}
\end{figure}

In General Relativity, the growth rate defined in Eqn. (\ref{eq:f}) is simply related to the growth of the gravitational potential $\Phi$. This relation is slightly modified in the coupled quintessence scenario, as we will now show. We begin by using Eqn. \eqref{eq:phi_mt} in the quasi-static regime (valid in the sub-horizon limit, deep inside the matter dominated epoch):
\be
\label{eq:phi_mt_2}
k^2\Phi = - 4\pi G a^2\left( \delta\rho_b + \delta\rho_{\rm c} + \delta\rho_{\rm de} \right).
\ee
It turns out that the dark energy perturbation is negligible compared to the contributions from the baryons and dark matter (we have checked this numerally). Then, remembering that $\delta \rho_{\rm m} =  \delta\rho_b + \delta\rho_{\rm c}$ and using the background equations, we can derive 
the following equation for $\Phi$: 
\begin{equation} 
\frac{\dot\Phi}{\Phi} = - {\cal H} + \frac{\dot\delta_{\rm m}}{\delta_{\rm m}} + \frac{\dot x}{x}\frac{\rho_{\rm c}}{\rho_{\rm m}}. 
\end{equation}
Here, $x$ is the quantity defined in Eqn. (\ref{eq:beta}). If we define
\begin{equation}\label{{eq:f_eff}}
f_\Phi := \frac{{\rm d}\ln(a\Phi)}{{\rm d}\ln a} =  \frac{(a\Phi)^\cdot}{{\cal H}(a\Phi)},
\end{equation}
then we find 
\begin{equation}\label{eq:f_Phi}
f_\Phi = f + \frac{\rho_c}{\rho_{\rm m}}\frac{{\rm d} \ln x}{{\rm d} \ln a}. 
\end{equation}
We see that in the uncoupled case, for which $x = {\rm const} = 1$, $f_\Phi$ and $f$ coincide. We plot the behavior of $f_\Phi$ in Fig. (\ref{fig:f_eff}). Whereas $f$ characterizes the growth of the density contrast in matter, $f_\Phi$ characterizes the growth in the gravitational potential $\Phi$. 
For $f_{\Phi}$ now, the departure from the uncoupled case is significant, especially at very late times - well within the reach of current redshift galaxy surveys. Crucially, this new growth rate gives us a direct measure of the gravitational potential $\Phi$ which describes the shapes of gravity wells into which galaxies fall, and the CMB is lensed. 

Having diverged slightly, we now return to our central question: are conformal and disformal effects at all separable observationally at the perturbation level? Unfortunately, it would appear not. Just as for the effective equation of state, it seems that what $x$ is doing at very late times (when the coupling suppression has already been lifted) is what dictates how either growth rate, $f$ or $f_{\Phi}$, will behave: the disformal oscillations, characteristic of a newly introduced scale, die out long before the present day, along with any hopes of discerning between the conformal and disformal factors. Again, just as at the background level, the dynamics of late time growth are determined by the late time behavior of the function $x$ - a degeneracy between $C$ and $D$ - \emph{no matter how it comes about}. Whether the two coupling types can be separated, again, depends on the epoch. Earlier on, a distinction is manifest, but at the later stages of universe evolution, the two act together, and the distinction blurs. As before, `later' is defined by the scale $M_D$ relative to the evolving Hubble scale.  

\subsection{The power spectra}
\label{spectra}
\begin{figure}
\begin{center}
\scalebox{0.6}{\includegraphics{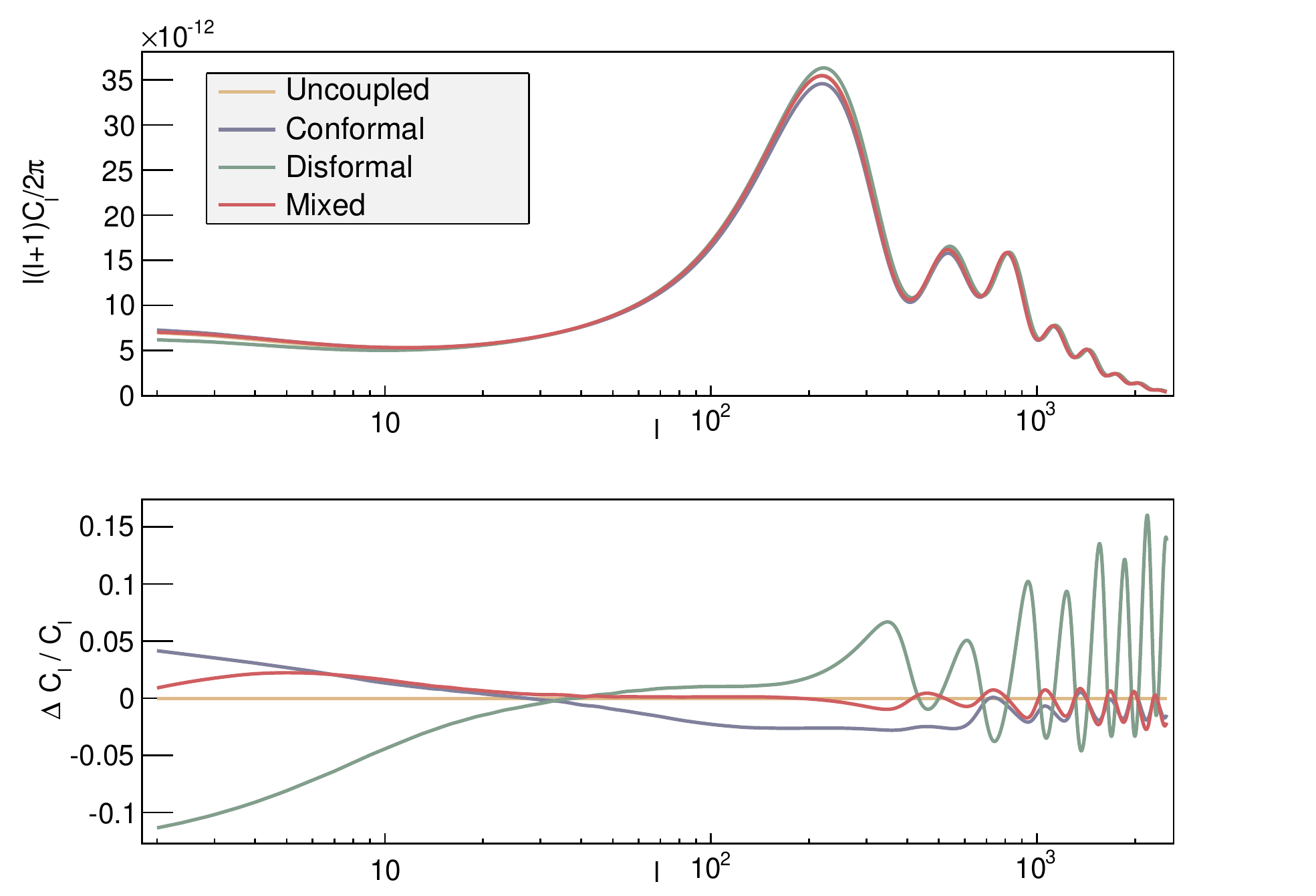}}
\caption{Top: angular power spectrum, $C_l$, against angular scale, $l$, for the four models described in table \ref{table:models} with parameters defined in equations \eqref{eq:free}. Bottom: fractional difference between the various models and the uncoupled case, i.e. $\Delta C_l/C_l:= (C_{l,\rm{i}}-C_{l,\rm{uncoupled}}) / C_{l,\rm{uncoupled}}$.}
\label{fig:Cl}
\end{center}
\end{figure}

We will now discuss the predicted CMB anisotropies and the matter power spectra. In Fig. \ref{fig:Cl} we show the angular power spectrum for the CMB anisotropies. For the type of models discussed in this paper, if a particular model has a reduction in power at low $l$ values it will have an enhancement at large ones, and vice versa, with respect to the uncoupled case. At small  multipole values the angular power spectrum is dominated by the integrated Sachs--Wolfe (ISW) effect, which is dependent on the late time behavior of the large scale gravitational potential. If a particular model undergoes enhanced expansion at latest times, quantified in an earlier section by an effective dark energy equation of state crossing the phantom line, then the gravitational potential $\Phi$ on large scales decays. The corresponding low $l$ anisotropies in the CMB are reduced as a consequence. For the same model the opposite happens for large multipoles: the anisotropies are enhanced. Since we have assumed fixed the boundary conditions for the cosmological parameter ($\Omega_{{\rm de},0} = 0.7$, ${\cal H}_0$, etc)  at the present time, the cosmological parameters differ (slightly) for the individual models at the time of decoupling. This results in different relative heights for the peaks at high multipoles. 

\begin{figure}
\begin{center}
\scalebox{0.6}{\includegraphics{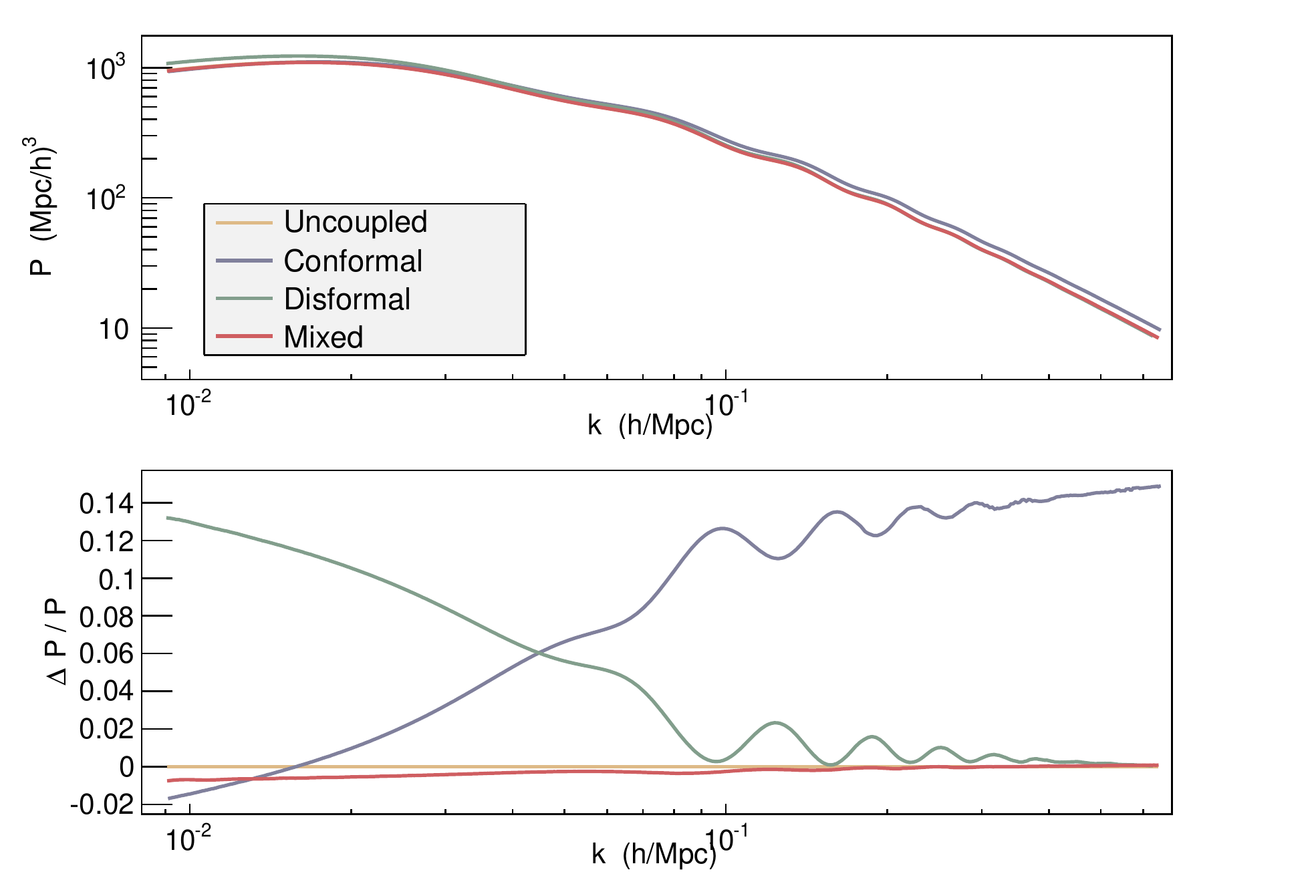}}
\caption{Top: matter power spectrum, $P$, against mode scale, $k$, for the four models described in table \ref{table:models} with parameters defined in equations \eqref{eq:free}. Bottom: fractional difference between the various models and the uncoupled case, i.e. $\Delta P/P:= (P_{\rm{i}}-P_{\rm{uncoupled}}) / P_{\rm{uncoupled}}$. h is the dimensionless reduced Hubble constant, defined as $H_0 = 100hMpc^{-1}.$.}
\label{fig:Pk}
\end{center}
\end{figure}

The matter power spectra are shown in Fig. \ref{fig:Pk}. For all models discussed here, there is an enhancement of power on small scales (large wave numbers), which is a direct result of the new scalar interaction between dark matter particles, which is always attractive. The peaks of the baryonic acoustic oscillations are shifted, which is due to fact that the cosmological parameters are different at time of decoupling. The couplings do not directly influence the position of the peaks. Strikingly, as for the $C_l$'s, the mixed model lies very close to the uncoupled case. As we found before, the conformal contribution to the effective coupling is suppressed by the disformal factor.

\subsection{Disformal Instabilities}
\label{sec:instabilities}
\begin{figure}
\begin{center}
\scalebox{0.6}{\includegraphics{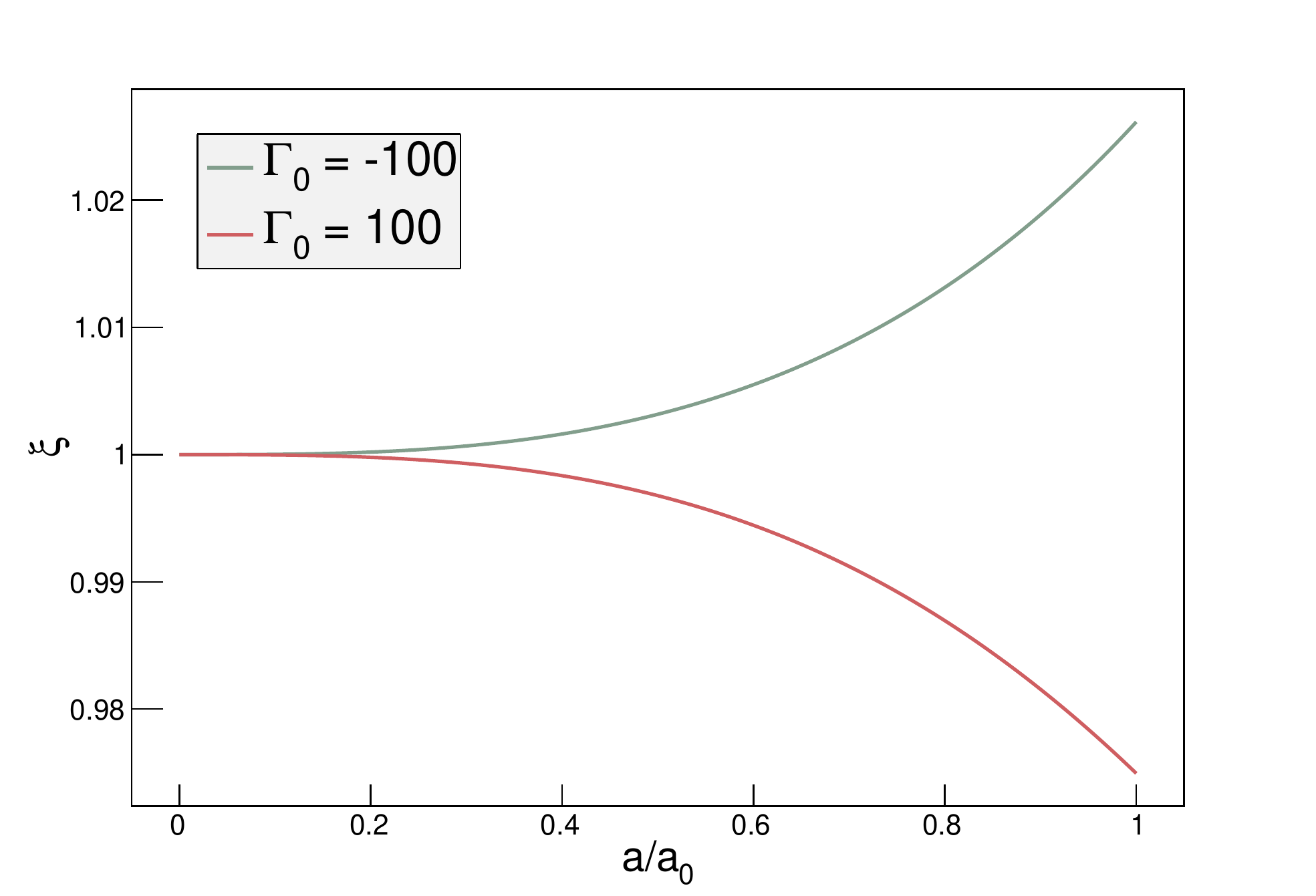}}
\caption{The instability function defined in \eqref{eq:xi} for two purely disformal models. For both curves, $\beta_V = -2$, $\beta_C=0$, $\beta_D=0$ and $M_D = M_V/\Gamma_0^{1/4}$. }
\label{fig:xi}
\end{center}
\end{figure}

Up to this point we have studied cases for which $D>0$. We will now discuss models in which $D$ is negative. As we will see, we find runaway growth of perturbations in the scalar field, which we feel justify our choice to neglect this case for the entire preceding study. To see how a disformal theory can become unstable in its perturbations, we can first re-write Eqn. \eqref{eq:KGperturbed} in a more suggestive form, for simplicity treating the case where $C = 1$, and $D$ is constant:
\be
\label{eq:KGperturbed_2}
\ddot{\delta \phi} + 2\mathcal{H}\left[ 1 - \left(\frac{3}{2} + \frac{\dot{x}/x}{\dot{a}/a}\right)\xi \right]\dot{\delta \phi} + (k^2 + a^2V'')[1-\xi]\delta \phi = S(a,k)
\ee
where, using that $|1+(\rho_cD)^{-1}| \gg |\frac{(\dot{\phi} / a)^2}{\rho_cD}|$, we have defined the parameter $\xi$ as 
\be
\label{eq:xi}
\xi := \frac{1}{1 + \frac{1}{\rho_cD}},
\ee
and all terms that do not contain $\delta \phi$ or its derivatives are collected in the source term $S(a,k)$. The homogeneous solution evolves according to 
\be
\label{eq:KGperturbed_3}
\ddot{\delta \phi}_a + 2\mathcal{H}\left[ 1 - \left(\frac{3}{2} + \frac{\dot{x}/x}{\dot{a}/a}\right)\xi \right]\dot{\delta \phi}_a + \omega^2 \delta \phi_a = 0,
\ee
with $\omega^2 : = (k^2 + a^2V'')[1-\xi]$. We see that $\omega^2$ is always positive if $D$ is positive, and will become negative if $D$ is negative. Eqn. \eqref{eq:KGperturbed_3} and Fig. \ref{fig:xi} demonstrate this clearly: for positive $D$, $\xi < 1$ and the effective oscillator frequency $\omega$ is real; for negative $D$, $\xi > 1$ and the frequency can become imaginary - the $\delta \phi_a$ solution becomes an exponentially growing function. So much for perturbations, but the background system is also unstable here. An epoch where $(\rho_cD)^{-1} \sim -1$, and hence the system traverses a pole, we can see is almost guaranteed to occur for negative $D$ at \emph{some} time $\tau$, whether this happens at higher redshift or in our future.

The above analysis pertained to a simplified model, however, what we have just shown encourages us to conclude that such an instability can occur wherever the disformal factor goes negative, and this is indeed what we find numerically.  Our results suggest that theories with negative disformal factors can not be considered as viable.  


\section{Conclusions}
In this paper we have studied the observational consequences of an extension to the coupled quintessence scenario, incorporating disformal terms. By keeping dark energy in the slow roll regime, we have focused not on solving the cosmological coincidence problem, but rather on searching for observable signatures of realistic (near $\Lambda$CDM) coupled dark sector theories. Studies like these are an imperative when so little about the dark sector is known and so little can be assumed. 

An important result of our study is that tension between a model with large conformal coupling ($\beta_C$ in our notation) and data can be alleviated by the addition of a disformal interaction. This is because the disformal contribution very effectively suppresses the coupling function $Q$ and it's linear perturbation $\delta Q$ for a significant portion of the universe's lifetime. The suppression is clearly manifest in both power spectra, as the predictions of the mixed model are very close to those of $\Lambda$CDM, although $\beta_C$ is of order one. We also find that, reformulating the theory in terms of a disformal scalar, the conservation equation for a coupled matter species is solvable, and we have used the solution to derive a condition indicating whether or not a given coupled dark energy model could be interpreted as exhibiting phantom behaviour. 

Our analysis of the perturbations tells us that, as in the standard coupled quintessence model, the disformal term does not affect the gravitational slip. Additionally the growth rate of the matter density contrast and the growth rate of the gravitational potential no longer coincide in models with couplings - we expect this will provide an observational key to breaking degeneracies between information contained in the CMB, gravitational lensing and LSS. We furthermore find that a negative disformal coupling $D$ generically induces dark energy instabilities: perturbations in the scalar field will eventually grow quasi-exponentially.

The results of this paper suggest that it is very difficult to discern between conformal and disformal effects when using only background observables and first order cosmological perturbations (e.g. CMB anisotropies or the matter power spectrum); an analysis of the non-linear regime on small scales it seems will be necessary to look for a `disformal smoking gun'. 

This preliminary study, far from complete, still must make direct contact with data. We leave this task for future work, where we intend to use CLASS's Monte Python to confront the expansive data set open to cosmologists today. It would also be interesting to investigate in how far the mixed model discussed in this paper can be mimicked by a time--varying conformal coupling, such as those studied in \cite{Baldi:2010vv}. Further, more of the theory's functional freedom ($\mathcal{L}_{\rm (DE)}$, $C$, and $D$) must be explored. How robust will our conclusions be under relocations within this function space? In addition, the question of the impact of quantum corrections needs to be addressed. Models with both conformal and disformal couplings between dark matter and dark energy are motivated from string theory (see \cite{Koivisto:2013fta}), and in these models, the scalar field is a DBI field where the functions $C$ and $D$ are specified by the extra--dimensional space. As a consequence, the effective coupling $Q$ has a different form and behaviour. We will turn our attention to such models in future work. 


\appendix
\section{Frame transformations: background}
\label{sec:frame_transformations}
We here derive the set of transformations relating background matter variables in the Jordan and Einstein frame of a generally coupled theory (see also \cite{vandeBruck:2013yxa} and for a discussion in considerable detail including perturbations see \cite{Minamitsuji:2014waa}). This will allow us to solve the background conservation equation for a coupled species in the Einstein frame. We begin with the definitions of the stress energy momentum tensors, and from there, using the map between them, compute the transformation rules.
\begin{subequations}
In the Jordan frame:
\be
\tilde{T}^{\mu\nu} := \frac{2}{\sqrt{-\tilde{g}}}\frac{\delta(\sqrt{-\tilde{g}}\tilde{\mathcal{L}}_{(DM)})}{\delta \tilde{g}_{\mu\nu}},
\ee
for which we can now impose a perfect fluid description, hence defining a Jordan frame energy density, $\tilde{\rho}$, velocity field $\tilde{u}^{\mu}$, and pressure, $\tilde{P}$:
\be
\label{eq:sem_jf}
\tilde{T}^{\mu\nu} = (\tilde{\rho}+\tilde{P})\tilde{u}^{\mu}\tilde{u}^{\nu} + \tilde{P}\tilde{g}_{\mu\nu}.
\ee
\end{subequations}
\begin{subequations}
In the Einstein frame:
\be\label{eq:sem_ef}
{T}^{\mu\nu} := \frac{2}{\sqrt{-{g}}}\frac{\delta(\sqrt{-\tilde{g}}\tilde{\mathcal{L}}_{(DM)})}{\delta {g}_{\mu\nu}},
\ee
where we then define:
\be
\label{eq:sem_ef}
T^{\mu\nu} = (\rho+P)u^{\mu}u^{\nu} + Pg^{\mu\nu}.
\ee
\end{subequations}
A map between the two objects can readily be derived \cite{Zumalacarregui:2012us}:
\begin{eqnarray}
\label{eq:map}
T^{\mu\nu} &=& \sqrt{\frac{\tilde{g}}{g}}\frac{\delta\tilde g_{\alpha\beta}}{\delta g_{\mu\nu}}\tilde{T}^{\alpha\beta}
\nonumber \\
&=& C^3\sqrt{1+\frac{D}{C}\phi,_{\mu}\phi\mathrm{'}^{\mu}}\tilde{T}^{\mu\nu}
\nonumber \\
&=& C^3\gamma\tilde{T}^{\mu\nu},
\end{eqnarray}
where we recognize the disformal scalar $\gamma$ from Eqn. \eqref{eq:s} which parameterizes the relative contribution of the disformal factor.
Note as $D\rightarrow0$, $\gamma\rightarrow1$.
Now, choosing the Einstein frame line element to be of Friedman Robertson Walker form:
\be
ds^2 = g_{\mu\nu}dx^{\mu}dx^{\nu} = a^2(\tau)[-d\tau^2 + \delta_{ij}dx^idx^j],
\ee 
means that, using Eqn. \eqref{eq:dismetric}, we get, in terms of the disformal scalar 
\be
\label{eqA:JFline}
d\tilde{s}^2 = \tilde{g}_{\mu\nu}dx^{\mu}dx^{\nu} = Ca^2(\tau)[-\gamma^2d\tau^2 + \delta_{ij}dx^idx^j].
\ee 

Given the background metric choice, the fluids are homogeneous, hence no forces are exerted between elements of the fluid - each element follows a geodesic dictated by the metric. This means the 4-velocity field can be computed directly from it: $\tilde{u}^{\mu}  = dx^{\mu}/|d\tilde{s}|$ and $u^{\mu}  = dx^{\mu}/|ds|$. Using this and the map \eqref{eq:map} we get the full list of variable transformations between the Jordan and Einstein frame background quantities:
\begin{subequations}
\label{eq:trans}
\begin{eqnarray}
\tilde{u}^{\mu} &=& \frac{1}{C^{1/2}\gamma}u^{\mu} = \frac{1}{C^{1/2}\gamma a}\delta^{\mu}_0 \\
\tilde{\Theta} &=& \frac{1}{C^{1/2}\gamma}\left[\Theta + \frac{3C,_{\phi}\psi}{2C}\right]\\
\tilde{\rho} &=& \frac{\gamma}{C^2}\rho \\
\tilde{P}    &=& \frac{1}{C^2\gamma}P \\
\tilde{w}    &=& \frac{1}{\gamma^2}w
\end{eqnarray}
\end{subequations}
where $\Theta = 3\frac{\dot{a}}{a^2} = 3H$, and $\psi:=\sqrt{2X}=\dot{\phi}/a$.

We are now in a position to solve the conservation equation for coupled matter in the Einstein frame. The Jordan frame stress tensor is conserved, as the matter it describes is uncoupled in this frame, so we can instantly write down:
\be
\label{eq:JFcons}
\tilde{\nabla}_{\mu} \tilde{T}^{\mu\nu} = 0,
\ee
where $\tilde{\nabla}_{\mu}$ is the covariant derivative metric compatible with $\tilde{g}_{\mu\nu}$. Given the transformations \eqref{eq:trans}, \eqref{eq:JFcons} reduces to:
\begin{eqnarray}
\label{eq:cons_back}
\dot{\rho} + a\left[ \Theta +  \frac{3C,_{\phi}\psi}{2C}\right]\left(\rho+\frac{P}{\gamma^2}\right) &=& 
	\partial_{0}\left(\mathrm{ln}\frac{\gamma}{C^2}\right)\rho,
\end{eqnarray}
which, as long as $\tilde{w}$ is constant, is exactly solvable:
\be
\rho \propto \frac{C^2}{\gamma}(C^{1/2}a)^{-3( 1 + w/\gamma^2 )}.
\ee
That $\tilde{w}$ be constant is not as restrictive a requirement as it sounds. In fact, as matter in the Jordan frame is uncoupled from the scalar, we expect $\tilde{w}$ to be constant wherever it is in $\Lambda$CDM. For example, one can show that for any relativistic species (photons, massless neutrinos. . .), $\tilde{w}=1/3$.

\section{Frame transformations: perturbations}
Derivation of the transformations between perturbation variables of the two frames proceeds in the exact same way, though this time we will not be able to solve the equations exactly, as it can not be done for the uncoupled case.

As discussed in section \ref{sec:perturbations}, $\delta$ will denote matter density contrast: $\delta := \frac{\delta \rho}{\rho}$, $\delta P$ the pressure perturbation, $\delta\phi$ is the perturbation of the scalar field and $\hat{\delta}$ a general perturbation operator. Then, working in the Newtonian gauge to first order:
\be\label{eq:pert_metric}
ds^2 = a^2(\tau)\left[ -(1+2\Psi)d\tau^2 + (1-2\Phi)\delta_{ij}dx^idx^j \right]
\ee
which means that:
\ba
\label{eq:pert_metric_dis}
d\tilde{s}^2 = Ca^2(\tau)[&-&(1+2A)\gamma^2d\tau^2 \nonumber\\
	&+& 2(\partial_iB)\gamma d\tau dx^i \nonumber\\
	&+& (1-2E)\delta_{ij}dx^idx^j]
\ea
where we recall the definitions of $A$, $B$, and $E$: 
\ba
A &=& \Psi + \frac{\hat{\delta} C}{2C} + \frac{\hat{\delta} \gamma}{\gamma} \\
B &=& \left(\frac{1}{\gamma}-\gamma\right)\frac{{\delta} \phi}{\phi'} \\
E &=& \Phi - \frac{\hat{\delta} C}{2C}~, 
\ea
where 
\begin{subequations}
\ba
\hat{\delta}\gamma &=& - \frac{D}{\gamma C}\left[ \left( \frac{D'}{D} - \frac{C'}{C} \right) X {\delta}\phi + \hat{\delta}X \right] \\
\hat{\delta} C &=& C' {\delta}\phi~.
\ea
\end{subequations}

Note that $\Psi-\Phi = A - E - \frac{\hat{\delta} C}{C} -\frac{\hat{\delta} \gamma}{\gamma} $. As a consequence, if in the Einstein frame $\Phi - \Psi = 0$, implying that the gravitational slip $\eta = \Phi/\Psi = 1$, the slip in the Jordan frame $\tilde\eta = E/A$ will, in general, depend on the coupling.   

Perturbations to the tensors given in \eqref{eq:sem_ef} and \eqref{eq:sem_jf} respectively gives:
\begin{subequations}
\ba
T^{\mu\nu}  &=&
(\delta\rho + \delta P)u^{\mu}u^{\nu} 
+ 2(\rho + P)\delta u^{(\mu}u^{\nu)}
+ \delta P g^{\mu\nu} + P\delta g^{\mu\nu} + P\Pi^{\mu\nu}
\\
\tilde{T}^{\mu\nu}  &=&
(\tilde{\delta\rho} + \tilde{\delta P})\tilde{u}^{\mu}\tilde{u}^{\nu} 
+ 2(\tilde{\rho} + \tilde{P})\delta \tilde{u}^{(\mu}\tilde{u}^{\nu)}
+ \tilde{\delta P} \tilde{g}^{\mu\nu} + \tilde{P}\tilde{\delta g}^{\mu\nu} + \tilde{P}\tilde{\Pi}^{\mu\nu},
\ea
\end{subequations}
where we have denoted the fluid's anisotropic stress $\Pi^{\mu\nu}$, which parameterizes higher moments of the fluid decomposition. And, just as before, by perturbing the map \eqref{eq:map} we can compute the transformations between perturbed matter variables:
\ba
\tilde{\delta} 
&=& \delta + -\frac{2\hat{\delta}C}{C} + \frac{\hat{\delta}\gamma}{\gamma} \\ 
(\tilde{\rho}+\tilde{P})\tilde{\theta} 
&=& \frac{1}{C^2}\left[(\rho+P)\theta + \left(\frac{P}{\gamma^2}-P\right)\theta_{\phi}\right] \\
\tilde{\delta P} 
&=& \frac{1}{C^2\gamma}\left[\delta P - \left(\frac{2\hat{\delta}C}{C} + \frac{\hat{\delta}\gamma}{\gamma}\right) P\right] \\
\tilde{\Pi}^{\mu \nu}
&=& \Pi^{\mu\nu}~.
\ea
These equations agree with \cite{Minamitsuji:2014waa}. $\theta$ is the velocity divergence field, defined as $\theta:=\partial_i(a\hat{\delta}u^i)$, and quantifies discrepancies between the fluid's velocity field and the underlying geodesic field. We have also defined for dark energy $\theta_{\phi} := \frac{k^2{\delta}\phi}{\phi'}$. Mathematically, every field permits a fluid description under a change of variables, and $\theta_{\phi}$ is the scalar's velocity divergence.

\acknowledgments We are grateful to M. Baldi, T. Koivisto, C. Llineares, D. Mota, N. Nunes and M. Zumalacarregui for useful discussions. We extend our gratitude to the authors of the well-written CLASS code, which was modified for our work. The work of CvdB is supported by the Lancaster- Manchester-Sheffield Consortium for Fundamental Physics under STFC Grant No. ST/L000520/1.

\end{document}